\documentclass[
 reprint,
 superscriptaddress,
 amsmath,amssymb,
aps
prb,
]{revtex4-2}

\usepackage[ruled,vlined,linesnumbered]{algorithm2e}

\usepackage{physics}
\usepackage[utf8]{inputenc}
 
\usepackage{hyperref}
\usepackage{graphicx}
\usepackage{rotating}

\usepackage{amsmath,amssymb,amsfonts,textcomp}
\usepackage{color}
\usepackage{xcolor}
\usepackage{float}

\usepackage[normalem]{ulem}
\usepackage{subcaption}

\usepackage[group-separator={,},group-minimum-digits=4]{siunitx}

\usepackage{xspace}

\newcommand{\vdbf}{{vDBF}}

\newcommand{\nrots}{{\tt n\_rots}}
\newcommand{\maxiter}{{\tt max\_iter}}
\newcommand{\thresh}{$\mathtt {\epsilon}$}
\newcommand{\convthresh}{{\tt conv\_thresh}}

\begin{document}


\title{Pauli propagation enables fast classical simulation of strongly correlated quantum systems}
\author{Chinmay Shrikhande}
\author{Arnab Bachhar}
\author{Jos\'e Aar\'on Rodr\'iguez-Jim\'enez }
\affiliation{Department of Chemistry, Indiana University,
Bloomington, IN 47405, USA}

\author{Edward F. Valeev}
\affiliation{Department of Chemistry, Virginia Tech, Blacksburg, VA 24061, USA}

\author{Nicholas J. Mayhall}
\email{nmayhall@iu.edu}
\affiliation{Department of Chemistry, Indiana University,
Bloomington, IN 47405, USA}



\begin{abstract}
Ground state energy estimation for strongly correlated quantum systems remains a central challenge in computational physics and chemistry. While tensor network methods like DMRG provide efficient solutions for one-dimensional systems, higher-dimensional problems remain difficult. Here we present a variational double bracket flow (vDBF) algorithm that leverages Pauli Propagation, a technique originally developed for classical simulation of quantum circuits, to efficiently approximate ground state energies. By combining greedy operator selection with coefficient-based fluctuation truncation and energy-variance extrapolation, we obtain sub-1\% relative errors for Heisenberg and Hubbard lattices of up to 128 qubits. For a 10$\times$10 Heisenberg lattice (100 qubits), vDBF obtains accurate results in approximately 2 minutes on a single CPU thread, compared to over 50 hours on 64 threads for DMRG. 
Accurate results are also obtained for systems that are difficult for quantum Monte Carlo: on the maximally frustrated $8\times8$ $J_1$-$J_2$ lattice and the hole-doped $8\times8$ Hubbard model, the tightest thresholds extrapolate to within  a few tenths of a percent of the best published variational estimates. We further test vDBF on the 84-qubit $\pi$-valence active space of hexabenzocoronene, where the most reliable extrapolated correlation energies agree with DMRG to within 1\%. These results demonstrate that classical simulation techniques developed in the context of quantum advantage benchmarking can provide practical tools for many-body physics.
\end{abstract}

\maketitle
\section{Introduction}\label{sec:introduction}
    The simulation of quantum many-body systems provides an indispensable tool for studying molecules and materials, often revealing features that are difficult or impossible to access experimentally. Improving our ability to perform such simulations is therefore one of the most important computational challenges in physics and chemistry, with direct implications for problems ranging from high-temperature superconductivity to heterogeneous catalysis. Unfortunately, strongly correlated systems remain extremely challenging, and no general-purpose classical method exists that can provide both reliable and efficient results for arbitrary systems.

    Most traditional techniques for approximating low-energy quantum systems fall into one of three categories: perturbative methods 
    (MPn~\cite{Mller1934NoteOA_mpn}, CCn~\cite{bartlett_manybody_1978,shavitt_many-body_2009,RevModPhys.79.291_cc,Bartlett1981ManyBodyPT_lcc_1}, 
    GW~\cite{GW_first_1965,marie2024gw_review,FAryasetiawan_1998_gw}, etc.), 
    stochastic methods (DMC~\cite{dmc_foulkes2001}, 
    FCI-QMC~\cite{booth_fmc,cleland2010communications_fciqmc}, 
    AF-QMC~\cite{motta2018ab_afqmc,qmc_random_walks_2003,PhysRevB.55.7464_afqmc}, etc.), or tensor network states 
    (DMRG~\cite{dmrg_garnet,dmrg_reiher,woutersDensityMatrixRenormalization2014,white_density_1992,white_ab_1999,white_density_2005,maDensitymatrixRenormalizationGroup2015a}, 
    PEPS~\cite{hyatt2019dmrg_peps,fishman2017faster_peps,ma2024approximate_peps,PhysRevB.97.045145_peps,vidal2007classical_peps,verstraete2004renormalization_peps}, 
    MERA~\cite{stoudenmire2018learning_mera,PhysRevLett.99.220405_mera}, 
    etc.). Each framework excels for certain classes of systems but has fundamental limitations. Perturbative methods are effective when mean-field treatments are qualitatively correct but break down when many-body interactions become strong. Stochastic methods can provide benchmark-quality results for ground state energies but are limited when the sign problem cannot be controlled~\cite{troyer2005computational}. Tensor network states provide ideal compactness for low-dimensional systems but become less efficient in higher dimensions. Despite the breadth of problems treatable by these and other methods, systems that combine strong correlation with high dimensionality remain challenging for classical simulation.

In light of rapid advances in quantum information science, quantum computation has become an appealing approach for such classically challenging problems. Quantum hardware continues to improve~\cite{google2025quantum, kimEvidenceUtilityQuantum2023, bluvstein2024logical}, 
with growing qubit counts and decreasing error rates, 
while algorithmic developments have reduced the quantum resources required for accurate simulations.  
However, current quantum hardware and algorithms are not yet capable of outperforming classical methods for many-body simulation, and the realization of an exponential speedup remains uncertain~\cite{leeEvaluatingEvidenceExponential2023}.

While quantum simulation has not yet demonstrated a definitive advantage over classical methods, its development has driven significant advances in classical simulation techniques. The 2023 IBM Eagle experiment~\cite{kimEvidenceUtilityQuantum2023}, which claimed evidence for quantum utility in simulating kicked Ising dynamics, sparked rapid development of classical methods that ultimately matched or exceeded the quantum results~\cite{begusicFastClassicalSimulation2023, begusicFastConvergedClassical2024,tindallEfficientTensorNetwork2024a}. Central to these classical responses were algorithms for approximating Heisenberg evolution of Pauli operators (Sparse Pauli Dynamics, Pauli Propagation, and related methods) which proved remarkably effective at simulating deep quantum circuits~\cite{rallSimulationQubitQuantum2019,
    begusicFastClassicalSimulation2023, 
    begusicFastConvergedClassical2024, 
    broers2025scalable,
    rudolphClassicalSurrogateSimulation2023a,
    loizeau2025quantum,
    muller2026pauliengine,
    schusterPolynomialtimeClassicalAlgorithm2024,
    cirstoiuFourierAnalysisFramework2024,
    fontanaClassicalSimulationsNoisy2025,
    begusicSimulatingQuantumCircuit2025,
    schusterPolynomialTimeClassicalAlgorithm2025,
    angrisaniClassicallyEstimatingObservables2025, angrisaniSimulatingQuantumCircuits2025, millerSimulationFermionicCircuits2025, robbiatiDoublebracketQuantumAlgorithms2025, rudolphPauliPropagationComputational2025,
    gonzalez-garciaPauliPathSimulations2025}. 
These tools were developed to benchmark quantum hardware, but their potential for classical ground state energy calculations remains largely unexplored. In this work, we combine Pauli Propagation with a variational double bracket flow (vDBF) approach for ground state energy estimation. Double bracket flows have recently emerged as a powerful framework in quantum computing \cite{gluza2024double}, serving as the foundation for algorithms such as Double-Bracket Quantum Imaginary Time Evolution (DB-QITE) \cite{gluza_ite}, optimized initializations for variational ground-state optimization \cite{robbiatiDoublebracketQuantumAlgorithms2025}, and a variety of other advanced quantum protocols~\cite{suzuki2025double, suzuki2025grover, alghadeer2025double}. 

By adapting this quantum-inspired framework for classical execution, we benchmark vDBF on the $10\times10$ antiferromagnetic Heisenberg model, the frustrated $8\times8$ $J_1$-$J_2$ lattice, the $8\times8$ Hubbard model at half-filling and at one-eighth hole doping, and the $\pi$-valence active space of hexabenzocoronene (HBC).


\section{Theory}\label{sec:theory}
    To simulate the results of a quantum computation, one generally aims to compute the expectation values of observables for states that have been evolved by a sequence of unitaries, 
    \begin{align}\label{eq:expval}
    \expval{O} = \bra{0}U_1^\dagger U_2^\dagger\cdots U_n^\dagger O U_n\cdots U_2 U_1\ket{0}.
    \end{align} 
    While this can be expressed equivalently in either the Schr\"odinger or Heisenberg picture, the choice to evolve the state, $\bra{\psi_n}O\ket{\psi_n}$, or the operator, $\bra{0}O_n\ket{0}$, or both~\cite{begusicFastConvergedClassical2024} presents distinct opportunities for approximations. 
    
    Due, in large part, to the fact that the wavefunction is generally a lower-rank object, the Schr\"odinger picture is typically preferable. 
    However, considerations of sparsity can often lean in favor of the Heisenberg picture. 
    Physically relevant Hamiltonians are generally sparse, while ground states are generally not~\footnote{While one can always find a change of basis that makes the ground state sparse (e.g., the eigenbasis), this is computationally intractable, 
    whereas Hamiltonians are generally sparse in computationally tractable bases (e.g., Pauli, or fermionic operator bases).}.
    Furthermore, when the state is no longer pure (i.e., noisy circuits or finite temperature), a density operator is needed, further reducing the computational distinction between the Heisenberg and Schr\"odinger pictures. 
    
    Choosing Pauli strings, $P_i$, (defined as the positive Hermitian tensor products of Pauli operators) 
    as an operator basis, we can identify several computational advantages for working in the Heisenberg picture
    arising from their extremely simple and well-characterized mathematical properties~\cite{aguilarFullClassificationPauli2024,kokcuFixedDepthHamiltonian2022, wiersemaClassificationDynamicalLie2024,magoulasCliffordTransformationsFermionic2025}.
    Paulis can be conveniently indexed in binary form, and operator multiplication can be efficiently evaluated 
    through simple binary operations on integers.
    Because any two Pauli strings either commute or anticommute, these can also be efficiently obtained through binary operations.
    Due to the involution property, Pauli strings have closed form exponentiation, making  evolution of a state,
        \begin{align}
        U_j(\theta)\ket{\psi}= e^{-i\tfrac{\theta}{2}P_j}\ket{\psi} = \cos(\tfrac{\theta}{2})\ket{\psi} -  i\sin(\tfrac{\theta}{2}) P_j\ket{\psi} 
        \end{align}
        or Pauli string, 
        \begin{align}\label{eq:pauli_evolution}
        U_j(\theta)^\dagger P_k U_j(\theta) = \nonumber\\ 
        \begin{cases} 
        \cos(\theta)P_k +  i\sin(\theta) P_jP_k & \text{if } [P_k,P_j]\neq 0 \\
        P_k& \text{if } [P_k,P_j] = 0
        \end{cases}
        \end{align}
    exceptionally convenient. 
    More general considerations of exact exponentiation have recently been explored by Izmaylov and coworkers~\cite{jayakumarFeasibilityExactUnitary2025} and Evangelista and Magoulas~\cite{evangelistaExactClosedformUnitary2025,magoulasSpinAdaptedFermionicUnitaries2025}.
    Furthermore, the Heisenberg picture Pauli evolution naturally exposes simplifications from Clifford or low-magic unitaries. 
    While a $\pi/2$ rotation about any Pauli string has the potential to create a maximally entangled state in the Schr\"odinger picture, this angle corresponds to a Clifford gate, which, by definition, maps Pauli strings to Pauli strings, leaving the coefficient structure (and thus sparsity pattern) of any Heisenberg operator unchanged.  
    For example, compare the action of a $\pi/2$ rotation on either a state with an operator:
    \begin{align}
        U_i(\tfrac{\pi}{2})\ket{\psi} &= \tfrac{1}{\sqrt{2}}\left(\ket{\psi} - i\ket{P_i\psi}\right)\\
        U_i(\tfrac{\pi}{2})^\dagger P_j U_i(\tfrac{\pi}{2}) &= P_k,
    \end{align}
    assuming $P_i$ and $P_j$ don't commute. 
    In the Schr\"odinger picture, this Clifford evolution of a state increases in complexity generally, whereas the Pauli string, $P_j$, is just converted to a different Pauli, $P_k$.  


\subsection{Pauli Evolution Truncation}
\begin{figure}
\includegraphics[width=.8\columnwidth]{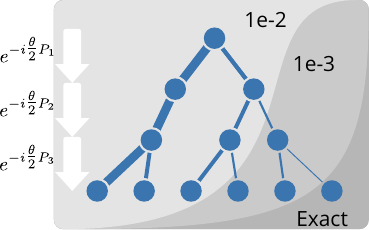}
\caption{Schematic depiction of the evolution of a single Pauli string (top) following 3 sequential small-angle Pauli rotations.
Background colors indicate the additional set of Paulis that are included under tighter levels of coefficient truncation.}
\label{fig:tree}
\end{figure}

Simulating the sequential rotation of an operator by a large number of Pauli strings, Eq.~\ref{eq:expval}, 
results in an exponentially increasing number of Paulis.
However, Eq.~\ref{eq:pauli_evolution} illustrates that this Pauli sum will potentially have a considerable amount of structure. 
Each rotation splits each non-commuting Pauli in the Pauli sum into two branches, a cosine and a sine branch, 
forming a binary tree. 
If each rotation angle is small, then the cos branches dominate, with the path weights diminishing exponentially in the number of sin branches.
In contrast, for near-Clifford rotations ($\theta \approx \pi/2$), the sine branch dominates. 
The schematic in Fig.~\ref{fig:tree}  illustrates the operator spreading that occurs during sequential evolution. 
Pauli evolution, therefore, creates a binary tree, such that the evaluation of the final expectation value requires one to sum the expectation values of all the leaves, weighted by the weight of the branch. 
While this is clearly exponential to do exactly, the fact that many of these paths will have extremely small weights leads to a number of truncation options. 

One could take a depth-first search approach to this, evaluating the sum path-by-path and aborting a given path once the current weight drops below a threshold. 
This requires essentially zero memory and is embarrassingly parallelizable; however, it performs rather poorly due to the need to go to very tight thresholds for convergence. 
One reason for this is that each path can visit the same Pauli, resulting in an opportunity for interference between the paths. 
Instead, we use a breadth-first search approach where the full operator at each time step is stored in memory, providing the opportunity to account for interference between Pauli paths. 
Keeping track of the full operator stabilizes truncation and facilitates the computation of properties such as 4-point correlation functions.
In this paper, we perform truncations after each Pauli rotation, deleting all Paulis whose coefficient absolute value is smaller than a given threshold $\epsilon$. 
This truncation approach is not new and has been used in several implementations, including Refs. \cite{rudolphPauliPropagationComputational2025} and \cite{begusicFastClassicalSimulation2023}.

The approximation of Pauli Heisenberg dynamics has been described with a variety of names,  Pauli Propagation (\texttt{PP})~\cite{rallSimulationQubitQuantum2019,rudolphPauliPropagationComputational2025}, Pauli Backpropagation~\cite{martinezEfficientSimulationParametrized2025}, Pauli Path simulation~\cite{gonzalez-garciaPauliPathSimulations2025}, and Sparse Pauli Dynamics (\texttt{SPD})~\cite{begusicFastClassicalSimulation2023}, 
each with different, but related goals. Although our current treatment of approximate Pauli evolution is most closely related to \texttt{SPD}, since we are currently only approximating the evolution by truncating small coefficient Paulis, we have also found truncations based on Pauli weight truncation to be effective, and so we will refer to our Pauli evolution treatment by the name of Pauli Propagation, 
referencing the first detailed presentation of the simulation strategy~\cite{rallSimulationQubitQuantum2019}.

\subsection{Double Bracket Flows}
Having access to an efficient algorithm for approximating unitary transformations has the potential to be useful for a variety of applications. 
The utility of quantum circuit simulation is clear and well-documented. 
In this work, we are interested in applying \texttt{PP} to the problem of ground state energy estimation.

In order to define a unitary transformation that evolves a state into the ground state, we start by adopting a double bracket flow formalism.
Double bracket flows are differential equations whose right-hand sides involve nested commutators, 
and which, when integrated, drive the system toward specific targets.
This approach was originally introduced as the Similarity Renormalization Group (SRG) in 1993 by G\l{}azek and Wilson,\cite{glazekPerturbativeRenormalizationGroup1994, glazekRenormalizationHamiltonians1993a} 
and was independently developed as a flow equation by Wegner in 1994.\cite{wegnerFlowequationsHamiltonians1994}

The basic idea is to define a Hamiltonian transformation along a continuous time-like parameter, $s$, that moves in a direction that minimizes some cost function. 
The ``canonical commutator''\cite{glazekPerturbativeRenormalizationGroup1994,glazekPerturbativeRenormalizationGroup1994,wegnerFlowequationsHamiltonians1994} is defined using the commutator of the Hamiltonian with the projection onto its own diagonal, $H_D$:
\begin{align}
    \frac{d}{ds}H(s) = [G(s),H(s)] = [[H_D(s),H(s)],H(s)].
\end{align}
Since $G(s)$ is an anti-Hermitian flow generator operator, this is analogous to a unitary evolution of $H$ by a time-dependent drive $G(s)$, 
a process that preserves the spectrum of $H(s)$ at all times, and is thus a similarity transformation.
This choice of $G(s)$ generates rotations that minimize the off-diagonal components to zero when the Hamiltonian commutes with its own diagonal. 
This form of $G(s)$ has been widely used, finding applications in physics~\cite{hergertInMediumSimilarityRenormalization2017}, chemistry~\cite{evangelistaDrivenSimilarityRenormalization2014, marieSimilarityRenormalizationGroup2023,wangAnalyticEnergyGradients2021},
 and quantum algorithms~\cite{gluza2024double, robbiatiDoublebracketQuantumAlgorithms2025}.

\subsubsection{State-specific flow generator}
While the ``canonical commutator'' would certainly suffice for obtaining our target eigenstate because it achieves full diagonalization,
it may be unnecessarily powerful, and we thus choose a flow generator that targets a single reference state instead,
$G(s) = [H(s), \rho]$. 
This generates evolution anytime the Hamiltonian does not have $\rho$ as a stationary state.
We choose $\rho$ to be the projector onto a reference computational basis state, which we will take, without loss of generality%
~\footnote{Any computational basis state can be rotated into $\ket{0}$, folding the resulting unitary into $H(0)$.}, 
to be the zero state, 
$\rho = \dyad{0}{0}$.
%
Since the reference state expectation value of the right-hand side is just the energy variance, which is non-negative,
the energy is guaranteed to decrease until $\ket{0}$ becomes an eigenstate:
\begin{align}\label{eq:eng_var}
    \frac{d}{ds}E(s) = -2(\expval{H^2}-\expval{H}^2)
\end{align}
We note that this choice of state-specific double bracket flow generator was rigorously derived in the DB-QITE framework~\cite{gluza_ite}, building on earlier DBQA work~\cite{gluza2024double}. In particular, Ref.~\cite{gluza_ite} provides the first systematic derivation of Eq.~\ref{eq:eng_var} and of the ansatz $[H(s), \rho]$ used here.
While this natural choice converges well, the projector expanded in the Pauli basis is highly non-sparse, requiring $2^N$ diagonal Pauli strings, 
\begin{align}\label{eq:proj_approx}
    \dyad{0} \propto I + \sum_i Z_i + \sum_{i<j} Z_iZ_j + \cdots.
\end{align}
To avoid this exponential cost, we choose to truncate this projector, 
where we approximate $\dyad{0}{0}$ by keeping only $Z$ strings up to a certain Pauli weight. 
For all results presented here, we keep only the 1-body terms.
While this might seem like a dramatic approximation, evolving by only the 1-body terms will still work, as this generator will only go to zero once $H(s)$ becomes block-diagonal, where Hamiltonian coupling between Hamming weight blocks are driven to zero. 
However, with this approximation, the integration of this equation takes longer to converge, as demonstrated in Fig. \ref{fig:n-body} of Appendix \ref{app:n-body}. Similar observations were also noted in Ref. \cite{xiaoyue2024strategies, robbiatiDoublebracketQuantumAlgorithms2025}.
Using this 1-body form, our double bracket flow generator becomes:
\begin{align}\label{eq:dbf_generator}
G(s) = \sum_i[H(s),Z_i],
\end{align}
leading to the following double bracket flow:
\begin{align}\label{eq:dbf}
    \frac{d}{ds}H(s) =  \sum_i[[H(s),Z_i], H(s)], 
\end{align}
which is identical to the form derived in Ref. \cite{robbiatiDoublebracketQuantumAlgorithms2025, gluza2024double}.

Integrating Eq. \ref{eq:dbf} for a small time, $\delta s$, we obtain the following recursive transformation:
\begin{align}
    H(s+\delta s) = e^{G(s)\delta s}H(s)e^{-G(s)\delta s}
\end{align}
Since our aim will be to approximate this process using \texttt{PP}, we will want to first expand $G(s)$ in the Pauli basis:
\begin{align}
 G(s) = \sum_ig_iP_i,  
\end{align}
where $g_i$ depend on $s$ and are purely imaginary. 
Since $\delta s$ is small, we can expand this solution in a product formula:
\begin{align}\label{eq:single}
    H(s+\delta s) = e^{g_nP_n\delta s}\cdots e^{g_1P_1\delta s}H(s)e^{-g_1P_1\delta s}\cdots e^{-g_nP_n\delta s}
\end{align}
The exact Riemannian-gradient flow is guaranteed to converge to the ground
state for almost every initial state with nonzero ground-state overlap: the
energy decreases monotonically, the critical points of the flow are exactly
the eigenstates of $H$, and all intermediate eigenstates are saddle points,
which first-order methods avoid with probability
one~\cite{pervez2025riemannian, malvetti2024randomized}.
In practice, however, a trajectory approaching a saddle point can
significantly slow convergence~\cite{zander2025role, wiersema2023optimizing,
pervez2025riemannian}.
Approximating this flow via sequential Pauli rotations (Eq.~\ref{eq:single})
with adaptive step sizes and a greedy selection of operators (detailed in the
next section) departs from the exact gradient direction; randomized variants
of such updates retain almost-sure convergence
guarantees~\cite{malvetti2024randomized, mcmahon2025equating}, and while these
guarantees do not transfer as a formal theorem to our deterministic, truncated
implementation, the fixed-point structure is preserved: each angle
optimization is energy-non-increasing, and the greedy gradients vanish only
when the reference has become an eigenstate of the evolved Hamiltonian.

While recent approaches have sought to implement these Riemannian gradient flows on quantum hardware by approximating the exact gradient via random projections into polynomial-sized Pauli subspaces \cite{pervez2025riemannian}, our deterministic vDBF approach allows for scalable classical simulation of much larger systems. Using the equation above, we \emph{could} proceed by using \texttt{PP} to sequentially evolve $H(s)$ by the series of Paulis in $G(s)$, 
each time truncating $H(s)$ by deleting Paulis with magnitudes smaller than $\epsilon$. 
However, keeping $\delta s$ small (a requirement to explicitly integrate Eq.~\ref{eq:dbf}) would require too many rotations for efficient convergence.
In the next section, we propose a modification of this strategy that accelerates this convergence.

\subsection{Pauli Propagation - Variational DBF (vDBF)}
At this point, we depart from the direct integration of the double bracket flow defined in Eq.~\ref{eq:dbf}. 
Rather than choosing the time steps to be ``small'', which would inevitably require many Pauli rotations to converge, 
we will instead choose an \emph{optimal} angle, $\theta^*_i$, for each Pauli, $P_i$, in the sequence, 
chosen to minimize the energy of the reference state, $\ket{\psi}$
\footnote{While this approach is valid for any computational basis reference state, the choice of Eq. \ref{eq:proj_approx} means that we must work with $\ket{0}$ as reference state by first transforming $H$
by a sequence of $X$ gates that transform $\ket{\psi}$ to $\ket{0}$, which is essentially a particle-hole transformation.} 
by analytically minimizing the cost function
\begin{align}
    F(\theta_i) &= \bra{\psi}e^{iP_i\theta_i }He^{-iP_i\theta_i}\ket{\psi}\nonumber\\
    &= \cos^2(\theta_i)\bra{\psi}H\ket{\psi}
    + \sin^2(\theta_i)\bra{\psi}P_iHP_i\ket{\psi} \notag \\[4pt]
    &\quad - 2i\cos(\theta_i)\sin(\theta_i)\bra{\psi}HP_i\ket{\psi}
\end{align}
This will ensure that each Pauli rotation minimizes the energy by the maximum amount possible. 
While this leads to a trajectory that deviates from the specific double bracket flow defined in Eq.~ \ref{eq:dbf}, 
we ultimately only care about the final state. 
Since we are no longer faithful to the flow of Eq.~\ref{eq:dbf}, the $s$ parameter is no longer meaningful, 
so we track the progress of this sequential transformation with an index, indicating the number of iterations:
$H(s) \rightarrow H^{(n)}$. 


\subsubsection{Greedy selection of generators}
Depending on the number of rotations performed during the vDBF, $H^{(n)}$ can accumulate many terms, 
resulting in a flow generator for the current iteration, $G^{(n)}=\sum_j[H^{(n)}, Z_j]$, that is even larger. 
However, not all Paulis in $G^{(n)}$ are equally important.
Ideally, we would only rotate by the most ``variationally helpful'' generators within each iteration to avoid unnecessary computation. 
To do this, we sort the operators in $G^{(n)}$ by the derivative of the energy with respect to rotation about each operator:
\begin{align}
    \frac{d}{d\theta_i}\expval{e^{i\theta_ig_iP_i}H^{(n)}e^{-i\theta_ig_iP_i}} 
    &= g_i\expval{[P_i,H^{(n)}]}\\
    =2g_i\Im{\bra{0}P_i\ket{\sigma^{(n)}}},
\end{align}
where $\ket{\sigma^{(n)}}=H^{(n)}\ket{0}$ is computed only once per iteration. 
This form allows the gradient vector to be efficiently to computed since we only need to read off the coefficient of the ket in $\ket{\sigma^{(n)}}$ whose {\tt on} bits match the {\tt X} string in $P_i$.

Having the operators in $G^{(n)}$ sorted by their reference state energy gradient, we then proceed to evolve by the first \nrots{} angles per iteration (user-specified simulation parameter). 
Setting \nrots $=1$ would therefore result in a single rotation by the operator with the largest gradient in the $G^{(n)}$. 
This would be a dynamic pool analog to how ADAPT-VQE~\cite{grimsleyAdaptiveVariationalAlgorithm2019,tangQubitADAPTVQEAdaptiveAlgorithm2021}
chooses operators out of a fixed operator pool. 
However, since evaluating the flow generator is not free, we find that it's generally more efficient to perform several rotations for each iteration, 
and so choosing \nrots{} to be tens to hundreds (e.g., 50-500) seems to work well. 
We note that this algorithm for designing a double bracket flow, which uses optimal steps to converge to the ground state, 
can be viewed as a classical version of the related quantum algorithm, Double Bracket Diagonal Operator Iterations~\cite{robbiatiDoublebracketQuantumAlgorithms2025}.

\subsubsection{Operator Truncation}\label{sec:truncation}
In order to tame the exponential growth of the Hamiltonian, 
we follow the coefficient truncation schemes used in both \texttt{SPD}~\cite{begusicFastClassicalSimulation2023,begusicFastConvergedClassical2024,begusicSimulatingQuantumCircuit2025} and PauliPropagation (\texttt{PP})~\cite{angrisaniClassicallyEstimatingObservables2025, angrisaniSimulatingQuantumCircuits2025, millerSimulationFermionicCircuits2025, rudolphPauliPropagationComputational2025},
by which an operator expressed as a linear combination of Paulis, $O=\sum_ic_iP_i$, 
is modified by deleting Paulis, $P_i$, that are considered unimportant. %
However, unlike the generic non-equilibrium dynamics of a quantum circuit, 
the vDBF approach has a well-defined reference state that allows us to significantly reduce truncation error, by truncating only the operator's reference state fluctuation instead of the full operator, 
\begin{align}
    P_i = \delta P_i + \bra{0}P_i\ket{0}I, 
\end{align}
where,  $\delta P_i=P_i- \bra{0}P_i\ket{0}I$.
In our simulations, whenever an Pauli's coefficient magnitude drops below a threshold,  $|c_i|<\epsilon$, we delete the operator's fluctuation: 
\begin{align}
    P_i\rightarrow \bra{0}P_i\ket{0}I.
\end{align}
Note that this truncation rule preserves the reference-state energy contribution of $P_i$ exactly, since $c_i\bra{0}P_i\ket{0}$ remains in H regardless of whether the operator's fluctuation is retained. The truncation therefore introduces no first-order error in the energy, with errors entering only through subsequent rotations of the truncated operator.
We refer to this truncation as $\textsc{clip}$ in Line 19 of the Algorithm \ref{alg:vDBF}. 
While Pauli weight~\cite{rudolphPauliPropagationComputational2025, schusterPolynomialTimeClassicalAlgorithm2025} (or Majorana weight for fermionic Hamiltonians)~\cite{millerSimulationFermionicCircuits2025}
can also be used to further decrease computational cost, we delay an in-depth analysis of this for future work (initial calculations seem to suggest it can be helpful)~
\footnote{We note that additional applications of the \textsc{clip} function are applied at various points to improve efficiency, 
but are not included in Algorithm \ref{alg:vDBF} for clarity. 
For example, after forming the flow generator, $G$, this operator is clipped. 
However, this is clipped with a tight threshold (default $10^{-6}$) so as to not affect the accuracy. }.

\begin{algorithm}
\caption{vDBF}
\label{alg:vDBF}
\KwIn{$H$, $\psi$, \thresh, \nrots, \maxiter, \convthresh}
\KwOut{$H^{(n)}$, $generators$, $angles$}
$generators \gets []$\;
$angles \gets []$\;
$H^{(1)} \gets H$\;
\For{$n \gets 1$ \KwTo~\maxiter}{
    \tcp{Compute generator}
    $G^{(n)} \gets [H^{(n)},\sum_jZ_j]$\;
    

    \BlankLine     
    \BlankLine 
    \tcp{Sort G by gradient}
    $gradients \gets []$\;
    $paulis \gets []$\;
    \For{$g_i,P_i \in G^{(n)}$}{
        Append $g_i\times\langle \psi | [P_i,H^{(n)}] | \psi \rangle$ to $gradients$\;
        Append $P_i$ to $paulis$\;
    }
    $perm \gets \mathrm{sortperm}(gradients)$\;
    $paulis \gets paulis[perm]$

    \BlankLine 
    \BlankLine 
    \tcp{Check for convergence}
    \If{$\mathrm{norm}(gradients) <$ \convthresh}{break\;} 

    \BlankLine 
    \BlankLine 
    \tcp{Rotate by optimal angles}
    \For{$j \gets 1$ \KwTo~ \nrots}{
        $P_j \gets paulis[j]$\;
        $\theta^*_j \gets \textsc{compute\_optimal\_angle}(H^{(n)}, \psi, P_j)$\;
        $H^{(n)} \gets \textsc{evolve}(H^{(n)}, P_j, \theta_j^*)$\;
        $H^{(n)} \gets \textsc{clip}(H^{(n)}, \epsilon)$\; 
        Append $\theta^*_j$ to $angles$\;
        Append $P_j$ to $generators$\;
    }
    $H^{(n+1)} \gets H^{(n)}$
}
\Return{$H^{(n)}$, $generators$, $angles$}
\end{algorithm}


\subsubsection{Zero Variance Extrapolation}\label{sec:extrap} 
Algorithm \ref{alg:vDBF} lists the steps required to converge a vDBF calculation for a given, user-specified accuracy control, $\epsilon$. 
Because convergence can be slow, iterating the algorithm until fully converged can become intractable for large systems. 
However, as one approaches convergence, a linear relationship between energy and variance ($\expval{H^2}-\expval{H}^2$) is expected. 
We leverage commonly used zero-variance extrapolations to estimate the energy at convergence, 
allowing us to stop iterating once a reliably linear relationship is observed.

\subsection{Comparison to existing methods}
The vDBF method is related to several other simulation techniques, both quantum and classical.
    \paragraph{ADAPT-VQE:} In ADAPT-VQE~\cite{grimsleyAdaptiveVariationalAlgorithm2019,tangQubitADAPTVQEAdaptiveAlgorithm2021}, one performs a variational minimization of an ansatz that has the form of a sequence of unitary transformations. The specific sequence of generators is defined iteratively such that at each iteration, the ground state energy gradient is used as an importance measure for determining which operator from a static, user defined operator pool should be added to the ansatz. After adding each new generator, all the operators are allowed to variationally relax. Our current vDBF algorithm can be viewed as a classical realization of a modified ADAPT-VQE, where only the last operator added is minimized (avoiding the challenging non-linear optimization). This step-wise parameter assignment shares conceptual similarities with the feedback-based quantum optimization algorithm FALQON \cite{magann2022feedback}, which also sets circuit parameters sequentially based on local gradient measurements rather than relying on global variational relaxation. However, unlike both ADAPT-VQE and FALQON, which typically draw from static, predefined operator pools, vDBF constructs its generators from a dynamically evolving operator pool. We also note that shortly after the first preprint of this paper was posted, and interesting related method called ADAPT-VMPE was published which combined ADAPT-VQE with \texttt{PP}~\cite{chakraborty2026scalablequantumcircuitgeneration}.
    
    \paragraph{iterative Qubit Coupled Cluster (iQCC):} vDBF can be viewed as a classical version of iQCC \cite{ryabinkinIterativeQubitCoupled2020, ryabinkinQubitCoupledclusterMethod2018} where the entanglers are chosen based on the flow generator in Eq.~\ref{eq:dbf_generator} instead of the direct interaction space, and the evolution is approximated using \texttt{PP}. Recent advancements also showed ways to reduce operator growth while lowering energy \cite{lang2023growth}, but vDBF relies solely on greedy steepest-descent selection. A recent preprint has also demonstrated a multistate variant, which showed promising results on small systems~\cite{langMultistateIterativeQubit2025}. 
     
    \paragraph{DB-DOI:} In the recently proposed quantum algorithm Double Bracket-Diagonal Operator Iteration (DB-DOI), the Hamiltonian is diagonalized through a double bracket flow on a quantum computer,\cite{robbiatiDoublebracketQuantumAlgorithms2025}
    where each sequential rotation can be determined variationally to minimize the off-diagonal elements of the Hamiltonian. 
    The current vDBF method can be viewed as a classical version of the DB-DOI approach, where the time steps are chosen to minimize the expectation values of the reference state, and the time evolution is approximated using the \texttt{PP} technique. 

\section{Numerical Simulations}\label{sec:results}
In order to assess the ability of vDBF, we first consider two canonical quantum many-body problems as examples:
an antiferromagnetic Heisenberg spin lattice and the Fermi Hubbard model. 
In both systems, we demonstrate the vDBF method's ability to provide fast estimates of ground-state energies. Finally, we extend our analysis to a complex molecular system, the polycyclic aromatic hydrocarbon hexabenzocoronene (HBC).

\paragraph{Extrapolation procedure}\label{sec:extrap_data}
To obtain stable extrapolations to the exact limit, we leverage the fact that when the trajectory is near convergence, we expect a linear relationship between the energy and the energy variance~\cite{PhysRevC.67.041301_variance_extrap_mizusaki2003,shimizu2010novel_extrapolation_monte_carlo_variance,shimizu2011extrapolation_monte_carlo_shell_model,variance_extrap_monte_carlo_mizusaki2012}.
However, further from convergence, the deviation from linearity might be non-negligible, 
and a simple linear regression might be less reliable despite having a seemingly respectable $R^2$ value. 
In order to obtain a more conservative estimate of errors in our extrapolations, 
we take the extrapolated result to be the average of a linear ($b_1$) and a quadratic ($b_2$) fit, 
with the uncertainties reported as half the difference. 
While we find that this extrapolation works surprisingly well, the fact that each calculation often generates thousands of data points means that one must decide how many data points to include in the extrapolation. 

In the calculations below, we start with the final step of a given optimization trajectory (which occurs due to either convergence or reaching the maximum number of iterations allowed), and then include all the previous points up to a cutoff which is chosen to minimize the uncertainty ($(b_1-b_2)/2$) and maximize the $R^2$ value of the linear fit. 
This balances precision with goodness of fit.
Using this ``optimal'' cutoff point in our data, our zero-variance extrapolations are reported as the average of the linear and quadratic fits, with the difference as the uncertainty: $E^\infty_{(v=0)} = \frac{b_1+b_2}{2} \pm \frac{b_1-b_2}{2}$.



All vDBF calculations were performed with our own software, {\tt DBF.jl}~\cite{nmayhallNmayhallDBFjl2025} and {\tt PauliOperators.jl}~\cite{nmayhallNmayhallPauliOperatorsjl2025}
which use standard symplectic Pauli representations for efficient binary manipulations. 
The DMRG calculations for Heisenberg and Hubbard models were carried out using the {\tt ITensors.jl} package \cite{fishmanITensorSoftwareLibrary2020}, and the HBC was done using {\tt BLOCK2} \cite{zhai2023block2}. SCI computations used the {\tt MPQC} package \cite{VRG:peng:2020:JCP}.

\begin{table*}
\begin{tabular}{l|rc|rcccc}
    \hline\hline
    Heisenberg &  \multicolumn{2}{c|}{DMRG} & \multicolumn{5}{c}{vDBF} \\
    Lattice  & & & $\epsilon:$ & {\tt 1e-3} & {\tt 1e-4}& {\tt 5e-5} & {\tt 1e-5}  \\\hline

    1$\times$100  & \bf Energy: &$-0.443230$ & \bf  Energy: & $-$0.439909 & $-$0.440248 & $-$0.440287 & $-$0.440336\\

    &&&\bf Extrap E: & $-$0.44120(3) & $-$0.44223(4) & $-$0.442363(2) & $-$0.4429(1) \\
    &&&&&&&\\
    & & &              Error:       & 0.459\% & 0.226\% & 0.196\% & 0.079\%\\
    & & &              Variance:    & 0.0020 & 0.0024 & 0.0024 & 0.0025\\
    & & &              Discarded W.:    & 0.054012 & 0.004627 & 0.002277 & 0.000376\\
    & &              & Iterations:  & 100         & 100      & 100        & 100 \\
    &  $\chi$: & 800 & len(H):      &  29272 & 249696 & 483163 & 2340704 \\   
        &  Time(min): & 3.9  & Time(min):   & 1.1 & 2.2 & 2.7 & 10.8\\
    &  Threads: & 64 & Threads:     & 1     & 1     & 1         & 1 \\\hline

    10$\times$10 & \bf Energy:& $-0.628455$ & \bf Energy:& $-$0.621682 & $-$0.621285 & $-$0.621338 & $-$0.621367 \\

        & \bf Extrap E: & $-0.628693$&\bf Extrap E: & $-$0.6277(2) & $-$0.627506(6) & $-$0.62780(3) & $-$0.6284(2)\\
    &&&&&&\\
    & &&              Error:               & 0.158\% & 0.189\% & 0.141\% & 0.045\% \\
    & &                   & Variance:      & 0.0145 & 0.0146 & 0.0147 & 0.0149\\
    & &                   & Discarded W.:  & 0.39230 & 0.06304 & 0.03624 & 0.00944\\
    & &              & Iterations:         & 100   & 100       & 100        &100 \\
    &  $\chi$: & 3500     & len(H):        & 189593 & 3023903 & 7159792 & 48357051
 \\
        &  Time(min): & 3213.9& Time(min):     & 2.5 & 18.2 & 39.4 & 249.8 \\

    &  Threads: & 64  & Threads:           & 1     & 1         & 1         & 1 \\\hline
    \hline
\end{tabular}
\caption{Energy/Site values for Heisenberg models for DMRG and vDBF.
vDBF calculations are each run with the threshold, \thresh{}, defined in the top row, 
and \maxiter=100 and \nrots=100. 
Energy denotes variational energies, while ``Extrap. E'' denotes extrapolated energies; note that while vDBF is variational in principle, coefficient truncation removes strict-upper bound in practice.  Uncertainties in the extrapolated vDBF energies, shown in parentheses, 
are given by half the difference between linear and quadratic extrapolation,
as described in Sec. \ref{sec:extrap_data}.
Discarded W. is the discarded weight from Sec. \ref{sec:extrap}.
1$\times$100 lattice has periodic boundary conditions.
and 10$\times$10 have open boundary conditions.
}\label{tbl:heis}
\end{table*}

\begin{figure*}
    \begin{subfigure}[t]{.5\linewidth}
        \centering
        \includegraphics[width=\linewidth]{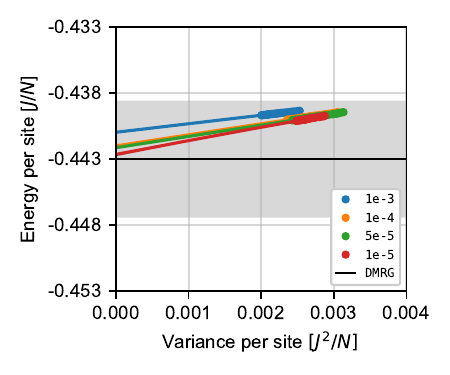}
        \caption{1$\times$100 Heisenberg Lattice}
    \end{subfigure}%
    \begin{subfigure}[t]{.5\linewidth}
        \centering
        \includegraphics[width=\linewidth]{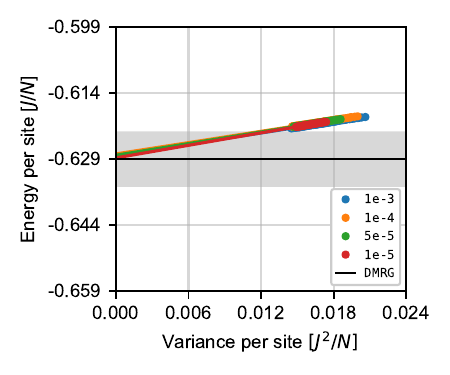}
        \caption{10$\times$10 Heisenberg Lattice}
    \end{subfigure}%
\caption{Energy vs Variance extrapolations for Heisenberg lattices using a series of vDBF thresholds: $\epsilon=$ {\tt 1e-3} (blue), {\tt 1e-4} (orange), {\tt 5e-5} (green), {\tt 1e-5} (red).
Extrapolations shown as solid lines.
DMRG best estimates (solid black line).
1\% error region w.r.t. DMRG best estimate (grey shaded area).}
\label{fig:heis_extrap}
\end{figure*}
\subsection{Heisenberg Hamiltonian}\label{sec:heisenberg}
The isotropic Heisenberg Hamiltonian is a standard model not only for describing magnetic interactions in materials but also for evaluating computational methods for strongly correlated quantum systems.
While the one-dimensional lattice admits an analytic solution via the Bethe ansatz, and DMRG provides an optimally efficient 
numerical treatment, the simulation in higher dimensions has no analytic solution, and precludes efficient simulation with a 1D MPS. 
While belief propagation algorithms for general tensor network state contractions have 
resulted in significant improvements recently~\cite{alkabetzTensorNetworksContraction2021b, evenblyLoopSeriesExpansions2024, tindallEfficientTensorNetwork2024a, tindallGaugingTensorNetworks2023a},
the description of higher dimensional strongly correlated systems is still challenging.

We start by computing the ground state energy of the antiferromagnetic Heisenberg spin Hamiltonian:
\begin{align}
    H = -2J\sum_{\expval{ij}}\vec{S}_i\cdot\vec{S}_j
\end{align}
on two lattices: (a) a large one-dimensional 1$\times$100 lattice, and (b) a large two-dimensional
10$\times$10 square lattice.
For the DMRG calculations, we used the {\tt ITensors.jl} software~\cite{fishmanITensorSoftwareLibrary2020}.
For 2D square lattices, we used snake pattern to define our MPO Hamiltonian and initial MPS, and ran symmetry number preserved calculations for half-filled state. 

For the vDBF results, we choose the N\'eel state for our reference state
$\ket{0101\dots}$. 
Because the form of the projector, $\rho$, and the resulting approximate flow generator (Eq. \ref{eq:dbf_generator}) assume the zero state, 
we simply transform our initial Hamiltonian by a sequence of $X$ gates so that we can work directly with the zero state:
\begin{align}
    E^{(0)} &= \bra{0101\dots}H^{(0)}\ket{0101\dots} \\
            &= \bra{0}\dots X_4X_2H^{(0)}X_2X_4\dots \ket{0}
\end{align}
Since each $X_i$ gate is a Clifford, the coefficient structure of $H^{(0)}$ is left unchanged. This feature will allow us to consider entangled reference states in future work. 

The results for the Heisenberg lattices are presented in Table \ref{tbl:heis}, 
and complementary convergence plots are provided in Fig. \ref{fig:heis_extrap}.

\subsubsection{1$\times$100 Heisenberg}\label{sec:1x100}
Looking first at the one-dimensional 1×100 Heisenberg lattice, we find that all 4 thresholds achieve less than 1\% error, 
with $\epsilon=${\tt 1e-3} doing so in just 1.1 minutes on a single thread.
We observe systematic accuracy improvement upon decreasing \thresh{}; 
each tighter calculation in the sequence decreases the deviations from DMRG, 
up to the \thresh{}={\tt 1e-5} result, which has approximately 0.08\% error. 
While both the energy errors and discarded weight decrease as \thresh{} is decreased, the cost also increases. 


This data already illustrates a more general feature that will be seen with the following examples as well: the accuracy obtained at a given threshold is controlled by a balance of two distinct error sources, (i) truncation error, which is directly reduced by decreasing $\epsilon$ (as seen in the discarded weight), and (ii) convergence error, which generally grows as $\epsilon$ is tightened, both because tighter thresholds empirically slow the convergence of the flow and because the increased per-iteration cost reduces the number of iterations affordable at fixed compute budget. However, since the energy-variance extrapolation lets us estimate the converged energy from a partially-converged trajectory, we still obtain accurate results from relatively few iterations.

In Panel (a) in Fig. \ref{fig:heis_extrap}, the convergence and extrapolations are shown for each of the four threshold values for the 1$\times$100 lattice. 
As described above, we choose our data cutoff point so as to minimize the uncertainty in our linear fit, and the shaded regions in the curve show the difference between linear and quadratic extrapolations.
For this system, the differences between linear and quadratic fits are undetectable in the plots, indicating a good fit. 
We find that even the raw energy, $\bra{0}H^{(n_\text{final})}\ket{0}$ without extrapolation, is already less than 1\% error from the ground state energy for all 4 truncation thresholds. 

\begin{figure}
\includegraphics[width=\columnwidth]{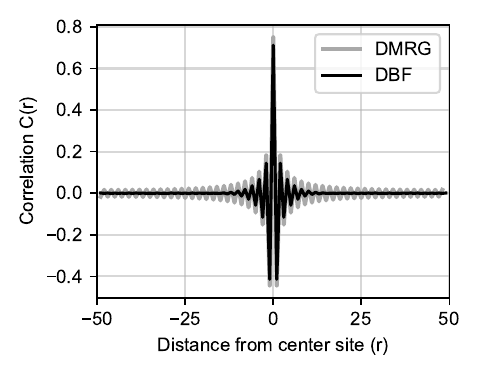}
\caption{Spin-spin correlation functions for 1x100 site Heisenberg lattice. $C(r) = \expval{S_1S_r} - \expval{S_1}\expval{S_r}.$ Blue are for converged DMRG results. Black line is \vdbf({\tt 1e-4})}
\label{fig:correlation}
\end{figure}

To develop better physical insight into the nature of our approximations, in Fig. \ref{fig:correlation} we have computed the spin-spin correlation function vs distance, $C(r) = \expval{S_1S_r} - \expval{S_1}\expval{S_r}$,
by simply transforming the target observable by the same sequence of unitaries that was constructed by the vDBF. 
This gives us raw expectation values, which are not expected to have the same accuracy as the energy, since we don't currently have a way to extrapolate observables to improve their accuracy, as we do with energy. 
However, this provides direct insight into the origin of the computational speedups. 
DMRG shows the expected polynomial decay of $C(r)$, with non-trivial alternating correlations extending across the entire 100-site chain.
In contrast, the vDBF correlations decay exponentially. Since the sequence of Pauli rotations is chosen to greedily minimize the energy, non-local correlations that don't directly affect the energy are naturally discarded. 
This also indicates that for gapped systems away from critical points (e.g., introducing anisotropy in $J$), we would expect vDBF to perform even better, since there the correlations also decay exponentially.  
Future work will need to focus on developing techniques for improving the accuracy of properties beyond just the ground state energy.

\subsubsection{10$\times$10 Heisenberg}
Compared to the previous example, the 10$\times$10 lattice is both large and high-dimensional, making direct TNS treatment challenging. 
In Table \ref{tbl:heis}, we obtained relatively good DMRG convergence with a bond dimension of 3500, which led to a very small extrapolation correction of around 0.0002 $J$/site.

All thresholds used for the vDBF calculations were able to obtain less than 1\% error with respect to the extrapolated DMRG energies, while requiring significantly fewer resources than DMRG. For example, the $\epsilon=${\tt 1e-3} calculation took only 2.5 minutes on a single thread, with an error of 0.158\%, whereas the DMRG calculation took more than 2 days on 64 threads.

 \begin{table*}
 \begin{tabular}{l|rc|rcccc}
     \hline\hline
     $J_1$--$J_2$ &  \multicolumn{2}{c|}{RBM+PP} & \multicolumn{5}{c}{vDBF} \\
     8$\times$8 Lattice  & & & $\epsilon:$& {\tt 8e-4} & {\tt 4e-4} & {\tt 2e-4} & {\tt 1e-4}  \\\hline
     $J_2/J_1=0.5$ & \bf Energy:& $-0.498964$ & \bf Energy:& $-0.487724$ & $-0.487900$ & $-0.487763$ & $-0.487671$ \\
     &&&\bf Extrap E: & $-$0.494272(1) & $-$0.493903(2) & $-$0.495942(7) & $-$0.497638(3)\\
     &&&&&&&\\
     & & &              Error:       & 0.940\%  & 1.014\%   & 0.606\%   & 0.266\% \\
     & & &              Variance:    & 0.0155   & 0.0149    & 0.0185    & 0.0218\\
     & & &              Discarded W.:& 15.341    & 11.327     & 7.152     & 4.382\\
     & &              & Iterations:  & 7278     & 5238      & 1632      & 798 \\
     &  & & len(H): & 4679162  & 14494361  & 40485851  & 106469502 \\
     &  &  & Time(min):   & 17.5   & 49.9      & 236.1  &  482.0 \\
     &  &  & Threads:     & 1     & 1         & 1     & 1 \\\hline
     \hline
 \end{tabular}
 \caption{Energy/Site values for the frustrated $8\times8$ $J_1$--$J_2$ model at $J_2/J_1=0.5$ (periodic boundary conditions) for vDBF
 compared against the RBM+PP reference energy~\cite{nomuraDiracTypeNodalSpin2021},
 the lowest V-score entry in the VarBench dataset~\cite{wuVariationalBenchmarksQuantum2024}.
 vDBF calculations are each run with the threshold, \thresh{}, defined in the top row and \nrots=50.
 Energy denotes variational energies, while ``Extrap. E'' denotes extrapolated energies; uncertainties, shown in parentheses,
 are given by half the difference between linear and quadratic extrapolation,
 as described in Sec. \ref{sec:extrap_data}.
 Error is measured on the extrapolated energy relative to the reference.
 Discarded W. is the discarded weight from Sec. \ref{sec:extrap}.
 }\label{tbl:j1j2}
 \end{table*}

\begin{figure*}
\centering
\begin{subfigure}[b]{.47\linewidth}
    \centering
    \includegraphics[width=\linewidth]{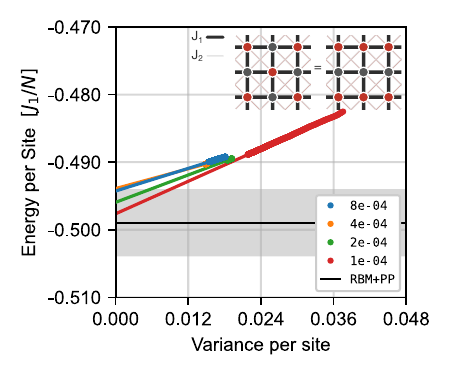} 
    \caption{Energy Extrapolations at classical critical point: $J_2/J_1=\tfrac{1}{2}$}
    \label{fig:j1j2b}
\end{subfigure}
\begin{subfigure}[b]{.52\linewidth}
    \centering
    \includegraphics[width=\linewidth]{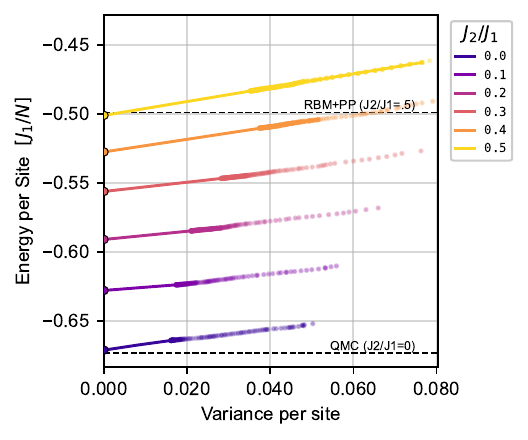} 
    \caption{Energy Extrapolations for range of $J_2/J_1$ ratios: $[0,\tfrac{1}{2}]$, $\epsilon=${\tt 8e-4}, $\maxiter=100$} 
    \label{fig:j1j2c}
\end{subfigure}
\caption{Frustrated Heisenberg 8$\times$8 lattice. (a) Energy vs Variance extrapolations for $J_2/J_1 = \tfrac{1}{2}$ using a series of vDBF thresholds: $\epsilon=$ {\tt 8e-4} (blue), {\tt 4e-4} (orange), {\tt 2e-4} (green), {\tt 1e-4} (red).
(b) Loose extrapolations ($\epsilon=${\tt 8e-4}, $\maxiter=100$) for range of $J_2/J_1$ ratios. 
         Extrapolations shown as solid lines.
         Best estimate (black line) for $J_2/J_1=\tfrac{1}{2}$ is RBM+PP obtained from Ref. \citenum{wuVariationalBenchmarksQuantum2024}.
         1\% error region w.r.t. best estimate (grey shaded area).
         Best estimate for $J_2/J_1=0$ from QMC.
      obtained from Ref.  \citenum{sandvikFinitesizeScalingGroundstate1997, sandvikHighprecisionGroundState2026}.
       Periodic boundary conditions.
        }\label{fig:j1j2}
\end{figure*}

\subsubsection{$J_1$-$J_2$ Heisenberg: Frustrated 8$\times$8 lattice}
The previous two examples focus on bipartite lattices,  which although they remain challenging to many simulation methods, are unique in that they do not suffer from a sign problem (since the rotation of one sublattice by local $\sigma_z$ makes all the off-diagonal elements non-positive).  Consequently,  the previous examples are able to be efficiently sampled using Monte Carlo techniques. In order to test whether or not the performance of \vdbf{} is somehow benefiting from this lack of sign-problem,
we consider a frustrated (non-bipartite) Heisenberg lattice by introducing next-nearest neighbor (diagonal) couplings, $J_2$. 
\begin{align}\label{eq:j1j2}
    H = J_1\sum_{\left<ij\right>}\vec{S}_i\cdot \vec{S}_j  + J_2\sum_{\left<\left<ij\right>\right>}\vec{S}_i\cdot \vec{S}_j,    
\end{align}
with periodic boundary conditions.
When $J_2=0$, the model reduces to simple Heisenberg model, and the N\'eel state is the lowest energy single configuration. 
When $J_2$ becomes large, a striped configuration becomes the lowest energy configuration, and the point at which these two phases cross ($J_2=J_1/2$) represents the maximal frustration point, where the two configurations are degenerate, see Fig. \ref{fig:j1j2b} (inset). 

While this degeneracy prevents one from selecting a ``good'' reference state for \vdbf, 
we can accelerate convergence by warm starting the \vdbf~ by first optimizing a sequence of rotations to minimize the reference state energy without truncation. 
In order to keep this untruncated warm-starting optimization from creating excessive operator growth, we choose the sequence of rotations to form a set of qubit-disjoint Pauli transformations. This sequence was defined by first coupling all horizontal edges with $XY$ Paulis, then vertical edges with $XY$ Paulis, then a 2$\times$2 tiling of $XXXY$ plaquette Paulis, e.g., 
\begin{align}
U(\theta) = e^{i\theta_5X_1X_2X_3Y_4}e^{i\theta_4Y_2X_4}e^{i\theta_3Y_1X_3}e^{i\theta_2Y_3X_4}e^{i\theta_1Y_1X_2}.
\end{align}
Since each 2$\times$2 plaquette is disjoint (equivalent to choosing a cluster mean field reference~\cite{jimenez-hoyosClusterbasedMeanfieldPerturbative2015,abrahamSelectedConfigurationInteraction2020c, mayhallUsingHigherOrderSingular2017a,abrahamClusterManybodyExpansion2021,bachhar_rocmf}), the operator spreading is minimal (H went from 768 terms to 15088 terms), making this optimization efficient such that no truncation was needed. 
This improved reference state lowered the initial energy from $-0.25~J_1$/site (N\'eel) to $-0.4375~J_1$/site (plaquette).%
\footnote{While this improved reference state made our calculations more efficient, we were still able to obtain sufficiently accurate results when starting from the original N\'eel state.}

In order to evaluate the performance of our \vdbf{} extrapolations, we compare directly to the Variational Benchmark dataset provided in Ref. \citenum{wuVariationalBenchmarksQuantum2024}.
From this dataset, we took $-0.49896348$, computed with RBM+PP, to be our best estimate reference number
as it had the lowest V-score for this system (0.00098). 

Considering the zero-variance extrapolations in Fig. \ref{fig:j1j2b}, we again observe similar behavior in that tightening the truncation threshold generally improves the extrapolated energy estimates, quickly bringing them to within the target 1\% of the best variational estimate from RBM+PP calculations~\cite{nomuraDiracTypeNodalSpin2021,wuVariationalBenchmarksQuantum2024}, 
with our tightest ({\tt 1e-4}) calculation providing an energy/site of $-0.4976$, as shown in Table \ref{tbl:j1j2}.
Although the $bare$ \vdbf{} calculations are far from converged (e.g., the {\tt 1e-4} calculation still had a variance/site of 0.02 after 798 iterations), 
the extrapolations are quite accurate. 
In fact, without extrapolation, the bare {\tt $\epsilon=$1e-4} \vdbf{} energy was comparable to the RBM($\alpha=1$)
energy from the same dataset, but with twice the variance. 

While computing the absolute ground state energy accurately is clearly desirable, often times, one is more interested in relative energies. 
In Fig. \ref{fig:j1j2c}, we show extrapolations from comparatively cheap calculations (100 iterations at $\epsilon=${\tt 8e-4} taking around 20 minutes total for all systems) on a series of $J_2/J_1$ ratios, ranging from 0 (the Heisenberg limit) to $\tfrac{1}{2}$ (the classical critical point). 
While we only have reference values for the end points, we observe quite systematic behavior allowing us to compare the extrapolated energies.
The computed energy difference between  $J_2/J_1=0$ and $\tfrac{1}{2}$ is only in error of about 2\%. 

While the \vdbf{} results in Table \ref{tbl:j1j2} show that we are indeed able to obtain sub-1\% accuracy for this spin-frustrated problem, these calculations were considerably more demanding than for the bipartite lattices considered in the previous sections. 
In Table \ref{tbl:frust_comp}, we show how both the number of terms in the Hamiltonian, as well as the final variance, increases as $J_2$ is increased from 0 to the maximally frustrated point $J_2=J_1/2$.
We observe that after 100 iterations, increasing $J_2$ results in a monotonic increase in the size of the resulting transformed Hamiltonian, with the critical point containing over 5$\times$ the number of Paulis than in the Heisenberg case. 
Further, the rate of convergence is also slowed for frustrated systems, with the variance at 100 iterations at $J_2=.5J_1$ over twice that of the Heisenberg case.
Nonetheless, vDBF still attains sub-1\% accuracy at the maximally frustrated point despite this nearly order-of-magnitude increase in operator count and the roughly twofold larger residual variance. This indicates that the accuracy of the method does not rely on the sign-problem-free structure of the bipartite examples, but rather on the energy-variance extrapolation, which seems to remain reliable even when the underlying trajectory is far from converged.
 \begin{table}[]
     \centering
     \begin{tabular}{l|ccc}\hline\hline
          $J_2$ & len(H) & Variance/site & Discarded W. \\\hline
          0          &    650,873 &   0.016 & 0.679 \\
          0.1        &    863,750 &   0.017 & 1.004 \\
          0.2        &  1,305,895 &   0.021 & 1.678 \\
          0.3        &  2,044,238 &   0.028 & 3.089 \\
          0.4        &  3,151,350 &   0.037 & 5.869 \\
          0.5 (critical point)    & 3,537,249 & 0.035 & 6.746\\
          \hline\hline
     \end{tabular}
     \caption{Effect of frustration on \vdbf{} efficiency after 100 iterations with \thresh{}={\tt 8e-4}.
     $J_1$ and $J_2$ are defined in Eq. \ref{eq:j1j2}.
     Both calculations run on 8$\times$8 lattice with periodic boundary conditions.}
     \label{tbl:frust_comp}
 \end{table}

\begin{figure*}
    \begin{subfigure}[t]{.5\linewidth}
        \centering
        \includegraphics[width=\linewidth]{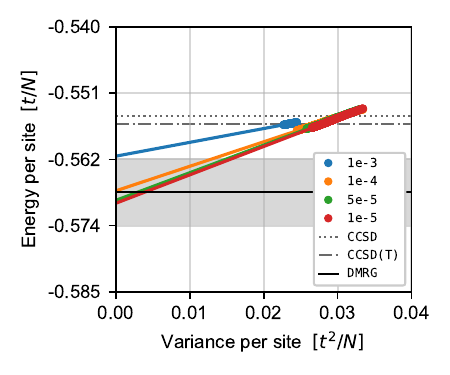}
        \caption{1$\times$64 Hubbard lattice}
    \end{subfigure}%
    \begin{subfigure}[t]{.5\linewidth}
        \centering
        \includegraphics[width=\linewidth]{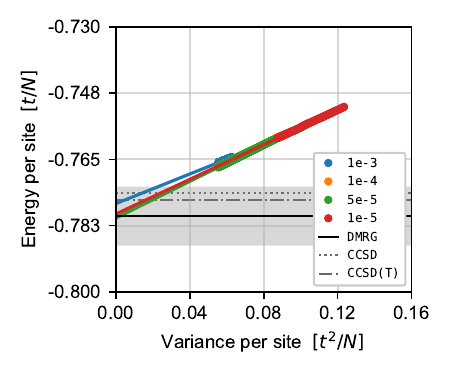}
        \caption{8$\times$8 Hubbard lattice}
    \end{subfigure}%
\caption{Energy vs Variance extrapolations for Hubbard lattice using a series of vDBF thresholds: $\epsilon=$ {\tt 1e-3} (blue), {\tt 1e-4} (orange), {\tt 5e-5} (green), {\tt 1e-5} (red).
Extrapolations shown as solid lines.
DMRG best estimates (solid black line).
1\% error region w.r.t. DMRG best estimate (grey shaded area).
}\label{fig:hubbard_extrap}
\end{figure*}


\begin{table*}
\begin{tabular}{l|rc|rc|rcccc}
    \hline\hline
    Hubbard   & \multicolumn{2}{c|}{DMRG} & \multicolumn{2}{c|}{SCI} & \multicolumn{5}{c}{vDBF} \\
    Lattice  & & & $\epsilon_1:$ & {\tt 2.5e-4}  & $\epsilon:$& {\tt 1e-3}& {\tt 1e-4}  & {\tt 5e-5} & {\tt 1e-5}  \\\hline
    
    1$\times$64   & \bf Energy:  &$-$0.567997 & \bf Energy: & $-$0.54671  & \bf Energy:  &$-$0.556624	&$-$0.557581	&$-$0.557266	&$-$0.557107\\

    &&&\bf Extrap:& $-$0.56426$^\dagger$ &  \bf Extrap: & $-$0.561953(1) & $-$0.56784(1) & $-$0.56945(1) & $-$0.5699547(7) \\
    &&&&  &  \\
    & &&&  &              Error:              &1.064\% & 0.028\% & $-$0.256\% & $-$0.345\%\\

    &  &                    & & & Variance:  &0.0227 & 0.0245 & 0.0258 & 0.0266\\
    &  &                    & & & Discarded W.:     & 0.7893 & 0.1294 & 0.0728 & 0.0183\\
    &  &                    & &  & Iterations:      &1000    &1000    & 1000	& 1000\\
    &  $\chi$: & 211       & $N_{det}$: & 13429409  & len(H):       &304797   & 4620991	   & 10418275			&69891297\\
    &  Time: & 6.6    & Time: & 26  & Time:   &38.1 & 227.8 & 451.5 & 3056.7\\
    &  Threads: &  64      & Threads: & 128  & Threads:       & 1      & 1      & 1      & 1 \\\hline
    
    8$\times$8   & \bf Energy: &$-$0.77272 & \bf Energy: & $-$0.76867  & \bf Energy:    &$-$0.766327	&$-$0.767390	&$-$0.767522	&$-$0.759833\\
    

    &\bf Extrap: &$-$0.7805(2) & \bf Extrap: & $-$0.77589$^\dagger$ &  \bf Extrap:& $-$0.777186(1) & $-$0.780655(3) & $-$0.78102(3) & $-$0.77995(4)  \\
    &&& & & & & & &\\
    & && & &               Error:              & 0.424\% & $-$0.020\% & $-$0.066\% & 0.070\%  \\ 

    &  &                    & & & Variance:    &0.0555 & 0.0555 & 0.0555 & 0.0873\\
    &  &                    & & & Discarded W.:      &1.6133 & 0.3268 & 0.2014 & 0.0277\\
    &  &                    &  & & Iterations:      &	1000    & 1000   &1000	&307\\
    &  $\chi$: & 6000       & $N_{det}$: & 9090855   & len(H):  &584905		&10622036			&26995422			&131216940 \\
    &  Time: & 5200.0  & Time: & 71 &   Time:    & 40.8 & 592.5 & 1462.9 &1693.4\\
    &  Threads: & 64        & Threads: & 128  & Threads:       & 1      & 1      & 1      & 1 \\\hline
    \hline
\end{tabular}

\caption{Energy/Site values for Hubbard models with 128 qubits. All lattices have open boundary conditions. Results compare DMRG, SCI, and vDBF methodologies. vDBF calculations are each run with the threshold, \thresh{}, defined in the top row, and \maxiter=1000 and \nrots=50. Uncertainties in the vDBF energies, shown in parentheses, are given by half the difference between linear and quadratic extrapolation, as described in Sec. \ref{sec:extrap_data}. The time reported are in minutes. $^\dagger$The $\epsilon_1=0$ SCI energy estimated via linear energy vs. variance relationship (similar to vDBF) using SCI energies with $\epsilon_1=\{10^{-3},5\times10^{-4},2.5\times10^{-4}\}$.}
\label{tbl:hubbard}
\end{table*}

\subsection{Hubbard Hamiltonian}\label{sec:hubbard}
The Fermi-Hubbard Hamiltonian:
\begin{align}
    H = -t\sum_{\ev{ij},\sigma}\hat{a}_{i\sigma}^\dagger\hat{a}_{j\sigma}
    + U\sum_{i}\hat{n}_{i\uparrow}\hat{n}_{i\downarrow} - \mu\sum_{i}(\hat{n}_{i\uparrow}+\hat{n}_{i\downarrow})
\end{align}
is a canonical model for describing interactions in electronic systems across a wide range of phenomenological settings. 
Changes in the balance of kinetic and potential energy ($t/U$) or connectivity $(\ev{ij})$ 
give rise to diverse physics, including metal-insulator transitions, magnetism, Mott insulating behavior, and high-temperature superconductivity.
Furthermore, it serves as a fundamental benchmark for evaluating new computational methods for many-body physics.\cite{simonscollaborationonthemany-electronproblemSolutionsTwoDimensionalHubbard2015}

In this section, we demonstrate vDBF's ability to estimate the ground-state energy of the Hubbard model. Having previously investigated the Heisenberg model, which governs the strong-correlation limit of the half-filled Hubbard model, we now focus on the intermediately correlated regime ($U = 4t$) at half-filling, with open boundary conditions. To target the half-filled ground state, we set the chemical potential within the Mott plateau ($\mu = U/2$) \cite{pavarini2016quantum}.
Analogous to the Heisenberg results, we first compute a large one-dimensional  1$\times$64 lattice (128 qubits) and then study a two-dimensional 8$\times$8 lattice (128 qubits) for which DMRG results were challenging to obtain, and validate these energies against Selected Configuration Interaction (SCI) calculations. We then turn to the hole-doped $8 \times 8$ lattice at U = 4t with periodic
boundaries, for which published reference energies
exist~\cite{levy2024automatic}. Because the half-filled clusters have open boundaries, their energies cannot be compared directly to the periodic lattice and thermodynamic-limit values of~\cite{simonscollaborationonthemany-electronproblemSolutionsTwoDimensionalHubbard2015}; we compare instead to DMRG and
SCI on the open boundary condition, and use the doped periodic lattice for the comparison against published benchmarks.

In order to simulate this fermionic system with this Pauli operator representation, we use a Jordan-Wigner transformation to map the fermionic sites onto distinguishable lattice sites. 
For the \vdbf{} calculations, we chose the N\'eel state as our reference state (alternating spin up and down).
For the DMRG calculations, similar to the 2D Heisenberg model, we used a snake-like mapping to define our site ordering.
The \vdbf, DMRG and SCI energies and convergence data are presented in Table \ref{tbl:hubbard}, 
and the extrapolation plots are shown in Fig. \ref{fig:hubbard_extrap}.

For the 1$\times$64 case, DMRG is well-converged at $\chi=211$ and no extrapolation is needed; vDBF errors are reported relative to the variational DMRG energy. For the 8$\times$8 case, DMRG requires bond-dimension extrapolation, and vDBF errors are reported relative to the extrapolated DMRG energy.

\begin{figure*}
    \begin{subfigure}[t]{.5\linewidth}
        \centering
        \includegraphics[width=\linewidth]{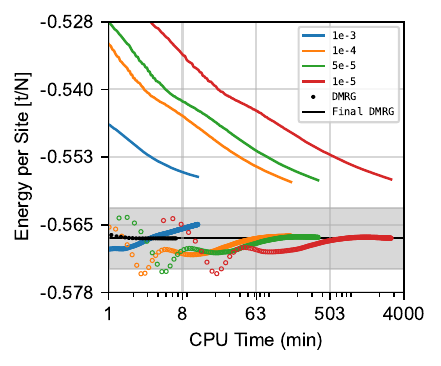}
        \caption{1$\times$64 Hubbard lattice}
    \end{subfigure}%
    \begin{subfigure}[t]{.5\linewidth}
        \centering
        \includegraphics[width=\linewidth]{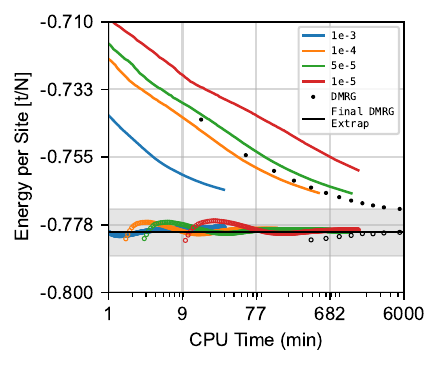}
        \caption{8$\times$8 Hubbard lattice}
    \end{subfigure}%
\caption{Energy estimates are plotted on a logarithmic time scale for the Hubbard lattices across four vDBF truncation thresholds: $\epsilon=$ {\tt 1e-3} (blue), {\tt 1e-4} (orange), {\tt 5e-5} (green), and {\tt 1e-5} (red). Solid lines indicate the raw variational trajectories, while hollow circles represent the corresponding linear energy-variance extrapolations computed using all prior iterations. The best extrapolated DMRG estimates are indicated by the solid black line, and the 1\% error threshold with the grey-shaded region.}
\label{fig:hubbard_time}
\end{figure*}

\begin{table*}
\centering
\begin{tabular}{l|cccc|rcccc}
   \hline\hline
   Doped & \multicolumn{4}{c|}{Reference energies} & \multicolumn{5}{c}{vDBF} \\
   $8\times 8$  & VAFQMC & VMC & FN& AFQMC & $\epsilon:$ & {\tt 1e-3} & {\tt 8e-4} & {\tt 6e-4} & {\tt 4e-4} \\\hline
   $U/t=4$ & $-1.013910 $ & $-1.00579$ & $-1.01153$ &$-1.01923$ &\textbf{Energy:} &$-$0.984450 & $-$0.984399 & $-$0.983505 & $-$0.982522 \\
   & & & & & \textbf{Extrap. E:} & $-$1.007577(6) & $-$1.011519(2) & $-$1.012138(4) & $-$1.013027(3) \\
   & & & & & & & & & \\
   & & & & & Error: &0.625\% & 0.236\% & 0.175\% & 0.087\%\\
   & & & & & Variance: & 0.1344 & 0.1392 & 0.1451 & 0.1517 \\
   & & & & & Discarded W.: & 7.242 & 6.489 & 5.576 & 4.474 \\
   & & & & & Iterations: & 1000 & 1000 & 1000 & 1000 \\
   & & & & & len(H): & 1852073  & 2556110  & 3873056  & 6961382  \\
   & & & & & Time: & 165.5 & 226.1 & 404.9 & 750.3 \\
   & & & & & Threads: & 1 & 1 & 1 & 1 \\\hline
   \hline
\end{tabular}
\caption{Doped Hubbard model on $8\times8$ lattice. Energy per site at $U/t=4$, $N_e=56$ ($n=0.875$), with periodic boundary conditions. vDBF calculations use the thresholds shown and $n_{\mathrm{rots}}=50$. ``Extrap. E'' is obtained from the energy vs. variance fit; uncertainties are half the difference between the linear and quadratic intercepts. Error is measured relative to the VAFQMC reference energy.}\label{tbl:hubbard_doped}
\end{table*}

\subsubsection{1$\times$64 Hubbard}

For the linear 1$\times$64 site Hubbard model, the extrapolated energy from {\tt 1e-4}, {\tt 5e-5}, and {\tt 1e-5} thresholds has an error of less than 1\% relative to the DMRG benchmark. As shown in Fig.~\ref{fig:hubbard_extrap}(a), our variational results surpass the accuracy of standard methods like Coupled Cluster Singles and Doubles (CCSD), and  Coupled Cluster Singles and Doubles with Triples correction CCSD(T). Using zero-variance extrapolations, the SCI calculations agree with the DMRG baseline, similar to vDBF, as shown in Table \ref{tbl:hubbard}. 


Consistent with the Heisenberg model, increasing accuracy by tightening the truncation thresholds naturally increases computational cost, as seen for {\tt 1e-5}, which took 3056.7 minutes, compared to {\tt 1e-3}, which took 38.1 minutes. 
However, zero-variance extrapolation dramatically lowers this computational cost. This can be seen from the Fig~\ref{fig:hubbard_time}(a), rather than targeting full convergence, the corresponding extrapolated energies (hollow circles) for the tighter thresholds drop into the sub-1\% error regime almost immediately, demonstrating the effectiveness of energy-variance extrapolations.
In fact, it took only 10 minutes to calculate a vDBF trajectory before the extrapolated error dropped to within 1\%.


\subsubsection{8$\times$8 Hubbard}
Shifting our focus now to the two-dimensional Hubbard model, we immediately note that in Table \ref{tbl:hubbard}, the DMRG results contain both a variational energy and an extrapolated energy, since we were not able to go beyond the bond dimension of 6000 while  
using a single compute node with 64 threads due to memory and compute time limitations.
As such, our DMRG extrapolation yields a substantial correction, amounting to approximately 1\% of the final energy estimate. Compared to DMRG, our SCI calculations with around 9 million determinants achieve a competitive extrapolated energy at a lower CPU time.

As we see in Fig~\ref{fig:hubbard_extrap}(b), all the thresholds obtain 99\% accuracy compared to the extrapolated DMRG results. 
We also obtain a better energy estimate compared to CCSD and CCSD(T).
Consistent with the Heisenberg model, the computational cost increases substantially as the truncation threshold is lowered. This is explicitly demonstrated by the evolved Hamiltonian at $\epsilon =$ {\tt 1e-5}, which grows to contain approximately 131 million Pauli terms while only reaching 307 iterations. We become memory-bound at this point, as the number of terms in the commutation calculation also reaches 600 million Pauli strings. 

As illustrated in Fig.~\ref{fig:hubbard_time}(b), the raw variational trajectories for both vDBF (solid lines) and DMRG (black points) converge slowly. In contrast, the variance-extrapolated vDBF energies (hollow circles) reach the sub-1\% error regime (grey shaded area) more quickly than the DMRG extrapolations, yielding highly accurate ground-state estimates after fewer iterations or less time. 
To compare them head-to-head, DMRG extrapolations took around 500 minutes to reach the 1\% region, whereas vDBF extrapolations fell to the 1\% error regime almost immediately.

\subsubsection{$8\times 8$ Doped Hubbard}

The doped two-dimensional Hubbard model is particularly difficult because it has the sign problem absent at half-filling. We therefore consider a periodic 8$\times$8 lattice with 56 electrons, equally divided between the two spin sectors, giving a density of $n=0.875$ and one-eighth hole doping. We use $U/t=4$, an intermediate-coupling regime where kinetic and interaction energies compete on equal footing, so a single-determinant description is unreliable. This provides a challenging test of vDBF convergence and energy-variance extrapolation.

For this example, we first converged a UHF solution from a checkerboard initial guess and used the UCCSD one-particle density matrix to construct natural orbitals. The occupied and virtual spaces were localized separately, producing the localized natural orbital (LNO) basis, which was used in the vDBF calculation after doing a Jordan-Wigner transformation. 
Although this orbital transformation increases the number of terms we start out with, the benefit from having a well-defined occupied-virtual separation ultimately provides faster convergence. 
To evaluate the doped Hubbard results, we compare the vDBF energies with four published reference values, as compiled in Table I of Ref. \citenum{levy2024automatic}: $-1.01153t$ per site from fixed-node (FN)~\cite{simonscollaborationonthemany-electronproblemSolutionsTwoDimensionalHubbard2015}, $-1.01391t$ from variational auxiliary-field quantum Monte Carlo (VAFQMC)~\cite{levy2024automatic}, $-1.00579t$ from Variational Monte Carlo(VMC)~\cite{wuVariationalBenchmarksQuantum2024}, and $-1.01923t$ from Auxiliary-Field quantum Monte Carlo (AFQMC)~\cite{shi2013symmetry,levy2024automatic}. We use the VAFQMC result as our main reference because it gives the lowest variational energy among these calculations.

\begin{figure}[H]
    \includegraphics[width=\linewidth]{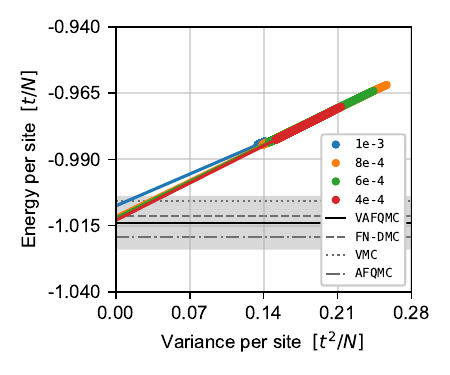}
    \caption{Energy vs Variance extrapolations for Doped $8\times 8$ Hubbard lattice using a series of vDBF thresholds: $\epsilon=$ {\tt 1e-3} (blue), {\tt 8e-4} (orange), {\tt 6e-4} (green), {\tt 4e-4} (red).
Extrapolations shown as solid lines.
VAFQMC best estimates (solid black line).
1\% error region w.r.t. VAFQMC best estimate (grey shaded area).}
\label{fig:dope_extrap}
\end{figure}

The results are shown in Fig.~\ref{fig:dope_extrap} and Table~\ref{tbl:hubbard_doped}. The reference values themselves are more scattered here than for the half-filled lattices: they differ from one another by up to 1.3\% (0.8\% among the variational methods alone). All four vDBF trajectories ran the full 1000 iterations. As $\epsilon$ is tightened, the extrapolated energy moves steadily toward the VAFQMC value, from 0.63\% above it at $\epsilon={\tt1e-3}$ to 0.18\% at $\epsilon={\tt6e-4}$ and 0.09\% at $\epsilon={\tt4e-4}$. Every extrapolation lies between the VMC and VAFQMC values, i.e., within the spread of the reference methods.

\begin{table*}
    \centering
    \setlength{\tabcolsep}{1pt}
    \begin{tabular}{rc|rc|rc|rcccc}
        \hline\hline
        \multicolumn{2}{c|}{DMRG} & \multicolumn{2}{c|}{TPSCI} &  \multicolumn{2}{c|}{SCI} & \multicolumn{5}{c}{vDBF} \\
           &   & & &$\epsilon_1:$ & {\tt 2.5e-5}& $\epsilon:$& {\tt 1.6e-4} & {\tt 8e-5} & {\tt 4e-5} & {\tt 2e-5}  \\\hline
        
        \bf Energy: &$-$0.424931 & \bf Energy: & $-$0.415978 & \bf Energy: & $-0.394326$ & \bf Energy:    &$-$0.405024 	&$-$0.4058708	&$-$0.404422	&$-$0.401284\\

        \bf Extrap: &$-$0.42653(2) & \bf Extrap: & $-$0.42618(7)  &\bf Extrap: & $-0.432597^\dagger$  &\bf Extrap: &$-$0.429585(5) & $-$0.42866(3) & $-$0.434184(6) & $-$0.437876(7) \\

        &&&&&&\\
         &&&& & &              Error:              &$-$0.716\% & $-$0.499\% & $-$1.794\% & $-$2.660\% \\
          &                    && & & & Variance:    & 0.0201 & 0.0188 & 0.0213 & 0.0248\\
          &             &&       & & & DW:      & 0.1120 & 0.0693 & 0.0440 & 0.0280\\
          &   & &                  & & & Iterations:      &	1000    & 1000   &1000	&1000\\
          $\chi$: & 1750       & &&$N_{det}:$ & 5965230 &  len(H):  &1717121			&4717996			&11120138		&25380248\\
        Time: & 1010.7  & &  &Time:& 37 &  Time:    &108.8 & 283.6 & 667.0 & 1608.3\\
          Threads: &64       & &  & Threads: & 128 &  Threads:       & 1      & 1      & 1      & 1 \\\hline
        \hline
    \end{tabular}

    \caption{Correlation Energy($E - E_\text{RHF}$) for HBC molecule.
vDBF calculations are each run with the threshold, \thresh{}, defined in the top row, 
and \maxiter=1000 and \nrots=50. 
Uncertainties in the vDBF energies, shown in parentheses, 
are given by half the difference between linear and quadratic extrapolation,
as described in Sec. \ref{sec:extrap_data}.
The percentage errors are calculated with respect to the DMRG correlation energy.
Timings for the TPSCI results are not compared, as the setup involved a complicated multi-step process, but the CPU cost is significantly higher than either DMRG or vDBF. $^\dagger$The $\epsilon_1=0$ SCI energy estimated via linear energy vs. variance relationship (similar to vDBF) using SCI energies with $\epsilon_1=\{2.5\times10^{-5},1\times10^{-5}\}$.
    }\label{tbl:hbc}
\end{table*}

\subsection{Molecular Hamiltonian - HBC $\pi$-system}\label{sec:hbc}
To evaluate vDBF's performance on strongly correlated chemical systems, we now target the ground-state energy of hexabenzocoronene (HBC), $C_{42}H_{18}$. As a large, symmetric polycyclic aromatic hydrocarbon, HBC exhibits non-trivial electron correlation, providing a rigorous molecular benchmark for our approach. 

The molecular Hamiltonian in second-quantized form is given as:
\begin{align*}
    H = \sum_{p, q}h_{pq}a^\dagger_p a_q + \sum_{p, q, r, s}g_{pqrs}a^\dagger_p a^\dagger_q a_s a_r
\end{align*}
where $h_{pq}$ and $g_{pqrs}$ are one-body and two-body electronic integrals. We first perform a restricted Hartree-Fock (RHF) calculation using the cc-pVDZ basis set, which yields a mean-field energy of $-1601.26918095$~Ha across a total of 678 spatial orbitals. To make the problem computationally tractable, we restrict the system to an active space of 42 electrons in 42 spatial orbitals, corresponding to the full $\pi$-valence space of the molecule. The active-space Hamiltonian is subsequently transformed into a localized natural orbital basis obtained from a CCSD calculation.

The fermionic creation and annihilation operators of this active space are then mapped to a Pauli string representation using the Jordan-Wigner transformation (84 qubits). To assess the accuracy of our vDBF correlation energies, along with DMRG, we also compare with the Tensor Product Selected Configuration Interaction (TPSCI)\cite{abraham2020selected,braunscheidelAccurateInterpretableRepresentation2024,braunscheidelGeneralizationTensorProduct2023a,bachharComputingExchangeCoupling2026} and SCI calculations.

\begin{figure}
    \includegraphics[width=\linewidth]{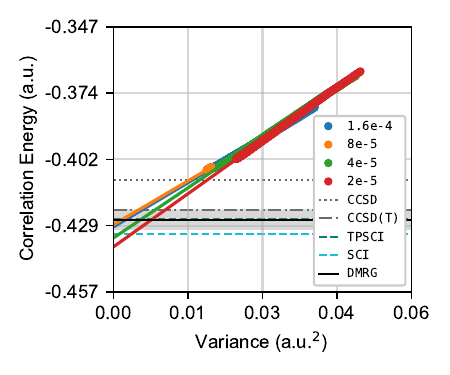}
    \caption{Correlation Energy vs Variance extrapolations for the HBC molecule using a series of vDBF thresholds: $\epsilon=$ {\tt 1.6e-4} (blue), {\tt 8e-5} (orange), {\tt 4e-5} (green), and {\tt 2e-5} (red). Extrapolations are shown as solid lines, with uncertainties (the difference between linear and quadratic fits) represented by the shaded regions surrounding them. The DMRG best estimate is indicated by the solid black line, while the 1\% correlation-energy error region relative to DMRG is shown as a grey shaded area. Reference energies are included for TPSCI (dashed teal line), SCI (dashed purple line), CCSD (dotted grey line), and CCSD(T) (dash-dot grey line).
}\label{fig:hbc_extrap}
\end{figure}

In Table~\ref{tbl:hbc} and Fig.~\ref{fig:hbc_extrap}, we summarize the ground-state correlation-energy calculations for the HBC molecule. As a benchmark, we performed DMRG calculations with a maximum bond dimension of $\chi=1750$, which was the largest value accessible within our node memory limits. The calculations are well-converged at this limit, as discarded weight was around $3\times10^{-5}$, and the extrapolated energy was only around 1.5 mH. 
To verify convergence, we also performed a high-level, computationally intensive TPSCI calculation, which was in close agreement with the DMRG estimate, suggesting that the DMRG/TPSCI values provide a reliable reference window for this active space. 
The SCI variance extrapolation overshoots the DMRG reference, recovering more correlation energy than the converged DMRG estimate, illustrating that variance extrapolation can fail differently for different methods.


The vDBF results for HBC follow a similar pattern as the lattice models, with one difference worth examining closely. The two loosest thresholds, {\tt1.6e-4} and {\tt8e-5}, extrapolate to within 1\% of the DMRG reference, at $- 0.72\%$ and $- 0.50\%$, respectively, while the two tightest, {\tt4e-5} and {\tt2e-5}, overshoot it by $1.79\%$ and $2.66\%$. 
This is initially surprising, as one would like ideally expect to see systematic improvement as one tightens the $\epsilon$ thresholds.

Investigating this non-monotonic convergence, we first note that the fidelity measure that directly depends on the truncation threshold, the discarded weight, does indeed behave as expected, systematically decreasing as we decrease $\epsilon$. 
However, the measure that depends on convergence of the flow (or optimization) directly tracks the accuracy here, decreasing from {\tt 16e-5} to {\tt 8e-5}, then increasing as $\epsilon$ is tightened. So this calculation seems to be limited, not by truncation, but rather flow convergence.
However, in principle, distance from convergence is not alone a good metric for determining the accuracy of the extrapolated energy. 
Once the energy/variance plot becomes truly linear, then the extrapolated energy is converged (although it is natural that the smaller the final variance is, the more linear we would expect the fit to be).
From this result, we see that the tighter thresholds take longer to reach the linear regime, and thus require many more iterations to converge the extrapolated energy. As such, we interpret the {\tt 8e-5} calculation to be the most robust HBC energy computed with \vdbf{}. 


To better understand the convergence behavior of the extrapolation (not just for HBC, but for all systems in this paper), in Appendix \ref{app:variance_grid}, we explore how the extrapolations behave as more iterations are included. 
Fig. \ref{fig:variance_all} plots both the energy as well as the extrapolated energy as a function of variance. Looking at the HBC curve (Fig. \ref{fig:variance_all}(g)), we see that 
while the loose threshold extrapolated energies are quite converged, the tighter calculations are still converging. From this plot is appears as though the tighter calculations {\tt 4e-5} and {\tt 2e-5} would approach the DMRG result with enough iterations.

\begin{figure}
    \includegraphics[width=\linewidth]{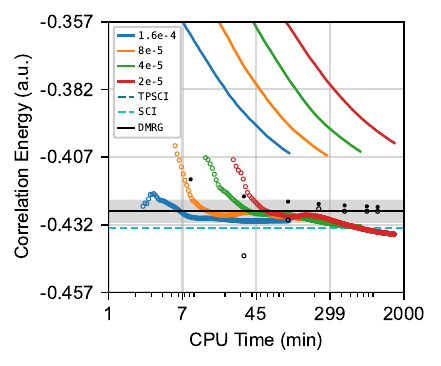}
    \caption{Correlation Energy vs Log Time for the HBC molecule using a series of vDBF thresholds: $\epsilon=$ {\tt 1.6e-4} (blue), {\tt 8e-5} (orange), {\tt 4e-5} (green), and {\tt 2e-5} (red). vDBF variational energies are shown as solid curves, while their corresponding linear extrapolations over time are shown as hollow circles. DMRG variational energies at different sweeps are marked by black dots, with the final DMRG extrapolated estimate shown as the solid black line. The 1\% correlation energy error region relative to the DMRG estimate is denoted by the grey-shaded area. Reference limits for TPSCI and SCI are shown as dashed teal and dashed purple lines, respectively.
}\label{fig:hbc_extrap_time}
\end{figure}

\section{Conclusions}

Classical simulation techniques developed to challenge claims of quantum utility, Sparse Pauli Dynamics or Pauli propagation, were originally designed to benchmark quantum hardware. Here, we have shown that the same ideas can also serve as efficient computational kernels for estimating ground-state energies of strongly correlated systems. Our variational double-bracket flow method, vDBF, combines Pauli propagation with greedy, gradient-based selection of Pauli generators, coefficient-based fluctuation truncation, and energy-variance extrapolation. This provides a compact operator-based route to ground-state energy estimation without explicitly storing a many-body wavefunction.


For the $10\times 10$ antiferromagnetic Heisenberg model (100 qubits), \vdbf{} reaches sub-1\% accuracy in approximately 2 minutes on a single CPU thread; the comparable DMRG calculation required over 50 hours on 64 threads. For the $8\times 8$ Hubbard model at half-filling (128 qubits), our DMRG calculations required $\chi = 6000$ and more than 80 hours over 64 threads, whereas \vdbf{} reaches this DMRG extrapolated energy to about 0.4\% in 40 minutes and 0.07\% or better at tighter thresholds. 

We also tested more challenging problems that exhibit a sign-problem. The maximally spin-frustrated $J_1$-$J_2$ model on an $8\times 8$  lattice extrapolates to within 0.3\% of the best variational reference, 
although the presence of spin-frustration results in a significantly greater number of Paulis that must be tracked. 
For the hole-doped $8\times 8$ lattice at $U/t=4$,  every \vdbf{} extrapolation considered fell within the target 1\% error and the tightest calculation had an error of only 0.09\% compared to the reference VAFQMC value. These results demonstrate that Pauli propagation can provide competitive ground-state energy estimates at substantially reduced computational cost for the lattice benchmarks considered here.

For a molecular example, we considered the full $\pi$-space of  hexabenzocoronene (84 qubits).
Here we found that the two loosest thresholds agree with our reference values to within 1\%, the best being $-0.50\%$ at $\epsilon=${\tt 8e-5}, while the two tightest overestimate the correlation energy due to the fact that the convergence is slower for tighter thresholds.  Overall, these results show that vDBF achieves competitive accuracy across both lattice and molecular systems.

The proposed vDBF method is primarily meant to serve as an energy-estimation approach as it can recover accurate ground-state energies even when the underlying truncated operator representation does not faithfully reproduce all ground-state observables. This limitation is visible in Fig. \ref{fig:correlation}, where the unextrapolated spin-spin correlations for the one-dimensional Heisenberg chain decay approximately exponentially rather than with the expected algebraic form. 

Several technical developments follow directly from these limitations. 
Further characterizing the convergence behavior as a function of thresholds, will be important for understanding the balance between operator fidelity and convergence criterion (as highlighted in the HBC results).
A second priority is enforcing physical symmetries more naturally. The current implementation uses individual Pauli-string rotations, which make it difficult to impose particle-number, spin, and other symmetry constraints exactly. While Appendix~\ref{app:prop_conserve} demonstrates that the resulting drift in both quantities over the course of a run is relatively modest, strict conservation is desirable for interpretation, and likely, computational efficiency. Efforts in this direction will follow the techniques developed for leveraging symmetry in the Pauli basis detailed in Refs. \citenum{haghshenasDigitalQuantumMagnetism2026} and \citenum{tengLeveragingSymmetryMerging2026}, we expect to see important property improvements using symmetry-preserving generators, grouped fermionic excitations, or explicit penalty terms and may improve both convergence and interpretability. 
A third priority will focus on developing a scalable parallel implementation following the work of Ref. \citenum{broers2025scalable} which will give us access to larger Pauli sums that can fit on a single node.
Finally, convergence at late iterations may be accelerated by limited non-local reoptimization, such as variationally relaxing the last k selected operators, or by coupling the Heisenberg-picture vDBF flow to a Schrödinger-picture correction inspired by Ref. \cite{begusicFastConvergedClassical2024}.

More broadly, this work shows that techniques developed to probe the boundary between classical and quantum computation can be redirected toward long-standing classical problems in strongly correlated matter. With improved extrapolation stability, symmetry control, and observable-specific extensions, Pauli-propagation-based methods may become useful tools for difficult problems ranging from frustrated magnetism and Hubbard-like models of correlated materials to extended $\pi$-electron systems and molecular active spaces relevant to chemical bonding and catalysis.

\section{Acknowledgments} 
We thank Marek Gluza, Zo\"e Holmes, and Manuel Rudolph for helpful comments and discussion.
The authors are grateful for the generous support from the  National Science Foundation (Award No. 2414574).
The work by EFV was supported by the U.S. Department of Energy via award DE-SC0022327.
This research was supported in part by Lilly Endowment, Inc., through its support for the Indiana University Pervasive Technology Institute.

\section*{Data Availability}

The software used to generate the vDBF results presented in this manuscript is publicly available. The vDBF algorithm is implemented in DBF.jl (\href{https://github.com/nmayhall/DBF.jl}{https://github.com/nmayhall/DBF.jl}), which depends on PauliOperators.jl (\href{https://github.com/nmayhall/PauliOperators.jl}{https://github.com/nmayhall/PauliOperators.jl}) for symplectic Pauli representations. DMRG calculations for the Heisenberg and Hubbard models were performed using the ITensors.jl package; DMRG calculations for hexabenzocoronene used BLOCK2; and selected configuration interaction calculations used MPQC. Input files, scripts, and numerical data needed to reproduce the figures and tables in this manuscript are available from the corresponding author upon reasonable request.

\bibliographystyle{achemso}
\bibliography{paper_library.bib, arnab.bib, chinmay.bib, ev.bib}
\newpage
\appendix
\section{Comparison to exact Projector}\label{app:n-body}
\begin{figure}
\includegraphics[width=\columnwidth]{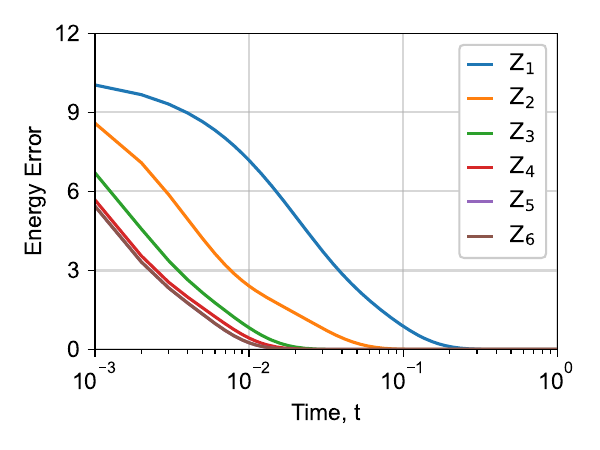}
\caption{Convergence of the exact continuous double bracket flow for different approximations to the projector for a  1$\times$6 Heisenberg lattice. 
$Z_i$ indicates that only terms up to $i$ are kept in the projector.}
\label{fig:n-body}
\end{figure}

In the theory section above, we motivated the use of the simple sum of single qubit $Z$ operators in the flow generator, $[H(s),\sum_iZ_i]$ as an approximation to the more complicated projector-based generator, $[H(s),\dyad{\psi}]$.
While this approximation was made for computational considerations, we show in Fig. \ref{fig:n-body} that this seemingly aggressive approximation still works to drive the system to the ground state. We plot the difference between the exact ground state energy, and the time-dependent expectation value of $H(s)$, for various approximations to the exact projection-based generator for a small system of a one-dimensional 6-site Heisenberg model. 
We see that while the full projector ($Z_6$) exhibits the fastest convergence ($Z_5$ is actually identical here, since $Z_6$ simply adds the $ZZZZZZ$ operator, which is a symmetry), all the other $n$-body approximations still eventually converge. 

While the $Z_1$ approximation is the slowest, the implementation is much more efficient. 
Further, in our final algorithm, the final step-size is not determined by directly integrating the double bracket flow, but rather chosen to variationally minimize the energy. 

\section{Truncation-corrected Variance}
In the plots above, we rely on zero-variance extrapolations to obtain accurate ground state energy estimates. 
As discussed in Sec. \ref{sec:truncation}, we explain how truncating only the energy fluctuation instead of the full operator provides a significant improvement to the accuracy to the energy. 
To see this, consider the Hamiltonian at iteration $n$ just before truncation:
$\tilde{H}^{(n)} = U^\dagger_nH^{(n-1)}U_n$, where $U_n = e^{-i\theta_nP_n}$. 
Truncation is performed by defining a projection (super)operator, $\mathcal{P}^{(n)}$, onto the Paulis in $\tilde{H}^{(n)}$ which are determined to be important, i.e., the Paulis which have coefficients $|c_i|>\epsilon$.  
The truncated operator, $H^{(n)}$, is then written in terms of this projector and its orthogonal complement, $\mathcal{Q}^{(n)}=1-\mathcal{P}^{(n)}$:
\begin{align}
\tilde{H}^{(n)} &= \mathcal{P}^{(n)}\tilde{H}^{(n)} + \mathcal{Q}^{(n)}\tilde{H}^{(n)}\\
H^{(n)} &= \mathcal{P}^{(n)}\tilde{H}^{(n)} + \expval{\mathcal{Q}^{(n)}\tilde{H}^{(n)}}.
\end{align}
Keeping the expectation value instead of just deleting the $\mathcal{Q}$ projection ensures that the operators expectation is unaffected by truncation, $\expval{H^{(n)}}=\expval{\tilde{H}^{(n)}}$, and the only approximation is then in the downstream dynamics.


While the interpretation of this as a fluctuation truncation is physically insightful, one could equivalently obtain this same result  in a more mechanistic approach by using ``bare truncation'' (where we don't add in the Q space expectation value) and simply computing the ``truncation error'' associated with projecting onto the $\mathcal{P}^{(n)}$ space, and accumulating this error over the sequence of rotations. 
For example, 
\begin{align}
    \Delta_E^{(n)} = \expval{\mathcal{P}^{(n)}\tilde{H}^{(n)}} - \expval{\tilde{H}^{(n)}} 
\end{align}
After $N$ rotations, the final energy would then be expressed as:
\begin{align}
    E^{(N)} =& \expval{\mathcal{P}^{(N)}\tilde{H}^{(N)}} - \sum_{n=1}^N\Delta_E^{(n)} \\
    =& \expval{H^{(N)}}
\end{align}

We similarly compute the final energy variance as:
\begin{align}
    V^{(N)} =& var\left(\mathcal{P}^{(n)}\tilde{H}^{(N)}\right) - \sum_{n=1}^N\Delta_{var}^{(n)},
\end{align}
where, $\Delta_{var}^{(n)}=var\left(\mathcal{P}^{(n)}\tilde{H}^{(n)}\right) - var\left(\tilde{H}^{(n)}\right)$.
This variance correction provides an estimate of how much of the variance reduction resulted from truncation, rather than convergence.

All the reported variances in this paper include this truncation correction.

\begin{figure*}
\includegraphics[width=\linewidth]{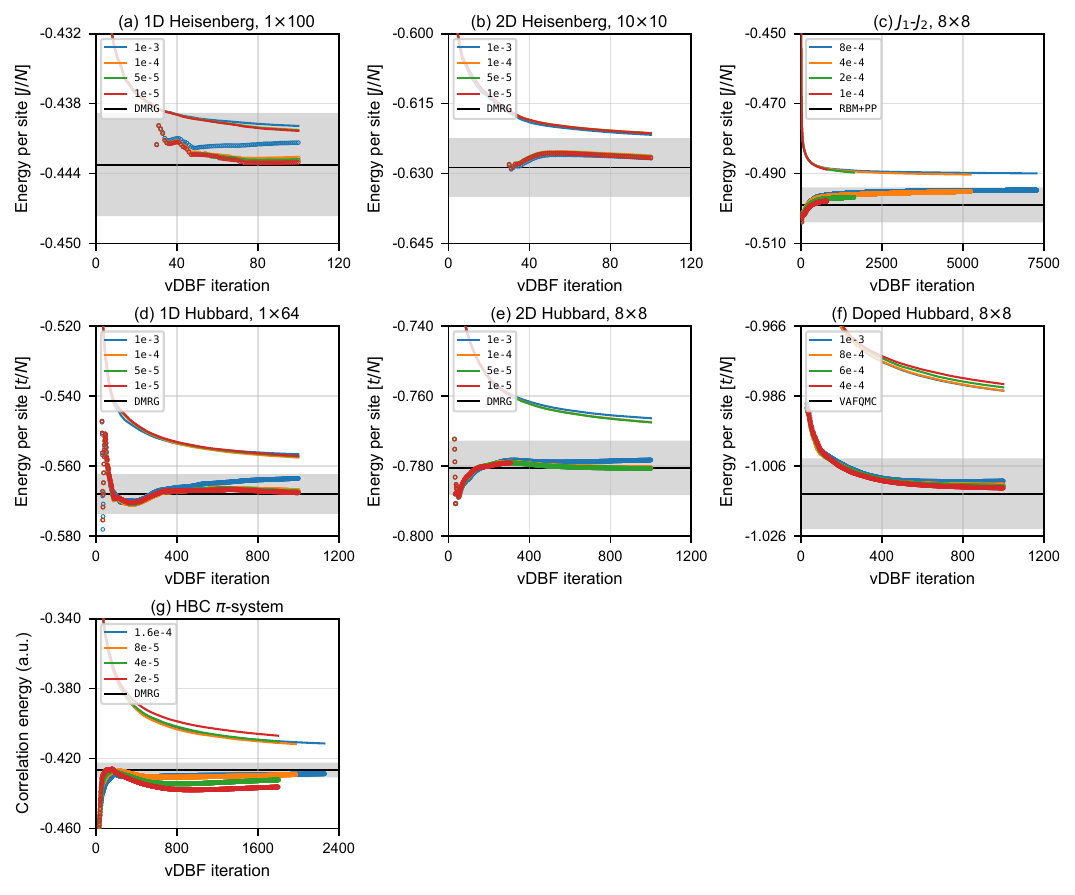}
\caption{Convergence of the zero-variance extrapolation across the benchmark systems. Solid curves are the raw \vdbf{} trajectories; open markers are the running intercepts for each iteration. For each raw data point ($E_i$, $V_i$), the extrapolation is performed using all the previous data points up to $2V_i$, ensuring that the extrapolation distance doesn't exceed the data range. Reference energies are the solid black lines, with the 1\% window shaded.}
\label{fig:variance_all_per_iter}
\end{figure*}

\begin{figure*}
\includegraphics[width=\linewidth]{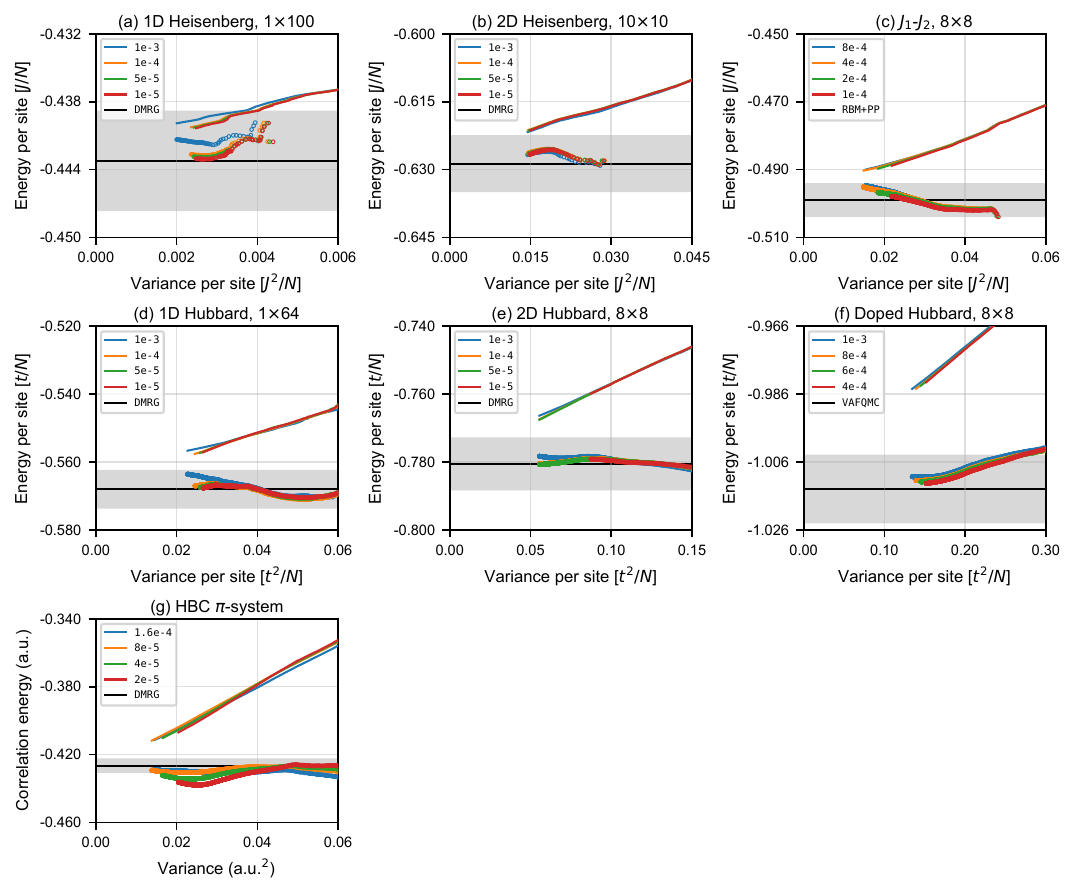}
\caption{Convergence of the zero-variance extrapolation across the benchmark systems. Solid curves are the raw \vdbf{} trajectories; open markers are the running intercepts, each drawn at the variance of its window's final iteration. For each raw data point ($E_i$, $V_i$), the extrapolation is performed using all the previous data points up to $2V_i$, ensuring that the extrapolation distance doesn't exceed the data range. Reference energies are the solid black lines, with the 1\% window shaded.
Same plot Fig. \ref{fig:variance_all_per_iter} but $x$-axis is variance instead of iteration.}
\label{fig:variance_all}
\end{figure*}

\section{Convergence of the Extrapolated Energy}\label{app:variance_grid}
Because \vdbf{} relies on extrapolation for quantitative accuracy, 
it's important to inspect the behavior and convergence of this extrapolation. 
In Figs. \ref{fig:variance_all_per_iter} and \ref{fig:variance_all}, we plot the convergence of both the raw energies, and the extrapolated energies as a function of iteration and variance, respectively.
In order to ensure a more interpretable extrapolation convergence plot, we choose the number of points in a slightly different manner than in the main text. 
Here, we perform the extrapolations with a rolling window such that $E_{extrap,i}$ is computed from the data points which have variance up to 2 times the variance at iteration $i$. 
Using this rolling $[V,2V]$ window ensures that the distance of extrapolation never exceeds the variance range of the data.

Considering each system in turn, we find that overall, 
the lattice systems generally display extrapolations that quickly converge within the 1\% error window. 
A couple exceptions can be noted. 
For both 1D Hubbard and Heisenberg, we see that the loosest threshold, {\tt 1e-3}, seems to exhibit significant truncation error which then results in an extrapolation trajectory that deviates from the tighter calculations. 
For the J1-J2 model, it appears as though the the extrapolations have not yet sufficiently converged, and will possibly increase in error with more iterations, although the tighter thresholds seem to be converging (with variance) a bit faster here.
However, for the molecular HBC example, we notice extrapolation behavior that's much further from converged for the tighter thresholds. 
While the looser {\tt 16e-5} and {\tt 8e-5} extrapolations appear to have flatlined within the 1\% region (consistent with Table \ref{tbl:hbc}), the tighter calculations are converging more slowly. 
It appears as though {\tt 4e-5} and {\tt 2e-5} would need significantly more iterations to converge their extrapolated energies to within the target error window. 

\begin{figure}
\includegraphics[width=\columnwidth]{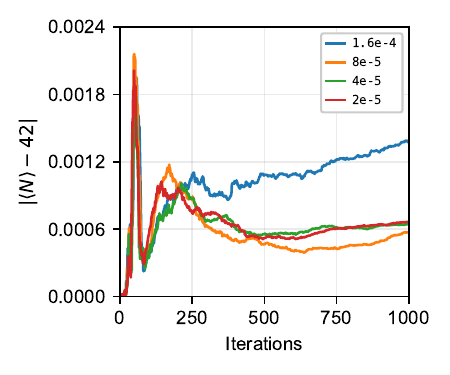}
\caption{Absolute error in the corrected expectation value of the evolved number operator with the circuits generated using different thresholds for the HBC molecule.}
\label{fig:hbc_elec}
\end{figure}

\begin{figure}
\includegraphics[width=\columnwidth]{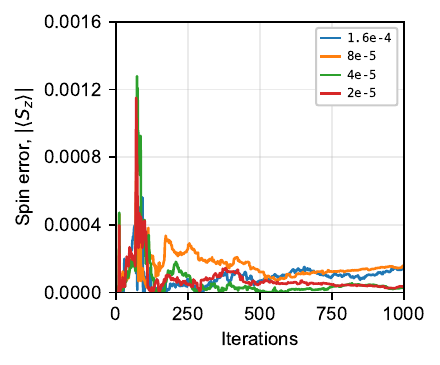}
\caption{Absolute error in the corrected expectation value of the evolved spin-projection operator with the circuits generated using different thresholds for the HBC molecule.}
\label{fig:hbc_sz}
\end{figure}
\section{Particle Number and Spin-Projection Conservation}
\label{app:prop_conserve}
In Sec. \ref{sec:1x100}, we observed limitations in the ability of \vdbf{} to capture long-distance spin-spin correlations in the Heisenberg example. For quantum chemistry applications, the ability to obtain accurate estimates of observables beyond just the energy is also valuable. In Figs. \ref{fig:hbc_elec} and \ref{fig:hbc_sz}, we plot the expectation values of the number and spin-projection operators as a function of iteration. Because individual Pauli rotations do not generally commute with either the number operator or the spin-projection operator, the average number of electrons and the spin projection are expected to fluctuate slightly over the course of a \vdbf{} optimization. From these plots, we do indeed see that neither the particle number nor the spin projection is constant, with deviations from the exact values (42 electrons and $S_z=0$, respectively) given on the $y$-axes.
Although particle number and spin projection are not exactly preserved, we do see that the deviations are quite small, generally in the 4th decimal place aside from brief early-iteration transients. The particle-number deviation is also biased toward a small net overall negative charge. In future work, we will explore adding either symmetry penalties or symmetry-constrained rotations to make the computation of properties more robust.

\section{Operator Weight Under the Jordan-Wigner Mapping}

\begin{figure*}
\includegraphics[width=\linewidth]{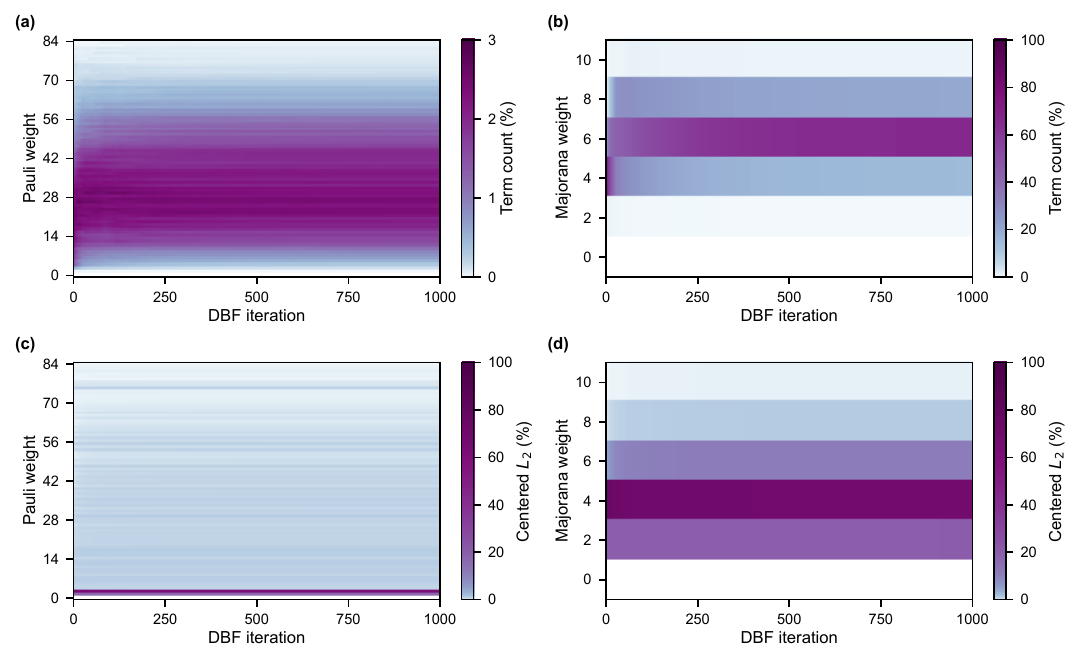}
\caption{Operator weight distribution along the HBC trajectory at $\epsilon = {\tt 2e-5}$ (84 qubits, 1000 iterations). Left panels are in the Pauli basis; right panels are in the Majorana basis. The top panels show the fraction of terms at each weight; the bottom panels show the fraction of the centered L2 norm. The identity contribution is removed from weight 0 in the lower panels.}
\label{fig:jw_weight}
\end{figure*}

The vDBF truncation acts on Pauli coefficients, and the Pauli weight depends on the fermion-to-qubit mapping. Under Jordan–Wigner, an operator acting on a few orbitals picks up a string of Z operators spanning the intervening qubits, so its Pauli weight reflects orbital indices as much as physical locality. The Majorana weight, twice the fermionic excitation rank, is independent of the mapping, so comparing the two separates the encoding from the physics.
Figure~\ref{fig:jw_weight} shows both for the HBC trajectory at $\epsilon = {\tt 2e-5}$. In the Pauli basis, panel (a), terms spread across weights 0 to 84 within the first hundred iterations, with most of the count between 14 and 56. In the Majorana basis, panel (b), the same operator has support only at weights 4, 6, and 8: the initial Hamiltonian is dominated by weight 4, a weight-6 component takes over during the first hundred iterations, and nothing above 8 acquires appreciable weight for the rest of the run. The apparent spreading is therefore a property of the encoding, not of the fermionic operator. What Jordan–Wigner renders as a weight-30 Pauli string is a two- or three-body fermionic operator.

The norm tells the same story. Panel (c) places almost all of the centered L2 weight at very low Pauli weight, and panel (d) at Majorana weights 2 to 6, with only a small contribution at 8 and above. Both distributions are stationary after the first hundred iterations, so the flow does not progressively drive the operator toward higher fermionic rank.
The truncation, therefore, does not discard high-body physics, and the results do not depend on the Jordan–Wigner ordering.

\end{document}